\documentclass[10pt, conference]{IEEEtran}

\usepackage{cite}
\usepackage{adjustbox}
\usepackage{graphicx,amssymb,lineno}
\usepackage{algorithm}
\usepackage{algorithmic}
\usepackage{multirow}
\usepackage[tight,footnotesize]{subfigure}
\usepackage{epsfig}
\usepackage{amssymb}
\usepackage{footmisc}
\usepackage{footnpag}
\usepackage{amsmath,paralist}
\usepackage{color}
\usepackage{perpage}
\usepackage{verbatim}

\IEEEoverridecommandlockouts

\begin{document}
\title{SDRBench: Scientific Data Reduction Benchmark for Lossy Compressors }

\author{\IEEEauthorblockN{ \hspace{-2mm}
Kai Zhao\IEEEauthorrefmark{3}, \hspace{-2mm}
Sheng Di\IEEEauthorrefmark{1}, \hspace{-2mm}
Xin Liang\IEEEauthorrefmark{2}, \hspace{-2mm}
Sihuan Li\IEEEauthorrefmark{3}, \hspace{-2mm}
Dingwen Tao\IEEEauthorrefmark{5}, \hspace{-2mm}
Julie Bessac\IEEEauthorrefmark{1}, \hspace{-2mm}
Zizhong Chen\IEEEauthorrefmark{3}, \hspace{-2mm}
and Franck Cappello\IEEEauthorrefmark{1}\IEEEauthorrefmark{4}
}

\IEEEauthorblockA{\IEEEauthorrefmark{1}
Argonne National Laboratory, IL, USA}

\IEEEauthorblockA{\IEEEauthorrefmark{3}
University of California, Riverside, CA, USA}

\IEEEauthorblockA{\IEEEauthorrefmark{2}
Oak Ridge National Laboratory, TN, USA}

\IEEEauthorblockA{\IEEEauthorrefmark{5}
Washington State University, WA, USA}

\IEEEauthorblockA{\IEEEauthorrefmark{4}
University of Illinois at Urbana-Champaign, IL, USA}

kzhao016@ucr.edu, sdi1@anl.gov, liangx@ornl.gov, \\sli049@ucr.edu, dingwen.tao@wsu.edu, jbessac@anl.gov, \\chen@cs.ucr.edu, cappello@mcs.anl.gov
% \vspace{-3mm}
\thanks{Corresponding author: Sheng Di, Mathematics and Computer Science Division, Argonne National Laboratory, 9700 Cass Avenue, Lemont, IL 60439, USA}
}

\maketitle
\thispagestyle{plain}
\pagestyle{plain}
\begin{abstract}
Efficient error-controlled lossy compressors are becoming critical to the success of today's large-scale scientific applications because of the ever-increasing volume of data produced by the applications.  
In the past decade, many lossless and lossy compressors have been developed with distinct design principles for different scientific datasets in largely diverse scientific domains. 
In order to support researchers and users assessing and comparing compressors in a fair and convenient way, we establish a standard compression assessment benchmark -- Scientific Data Reduction Benchmark (SDRBench)\footnote{Available at https://sdrbench.github.io}. SDRBench contains a vast variety of real-world scientific datasets across different domains, summarizes several critical compression quality evaluation metrics, and integrates many state-of-the-art lossy and lossless compressors. 
We demonstrate evaluation results using SDRBench and summarize six valuable takeaways that are helpful to the in-depth understanding of lossy compressors.  
\end{abstract}
\section{Introduction}
\label{sec:intro}

Today's high-performance computing (HPC) applications produce extremely large amounts of data, introducing serious storage challenges and I/O performance issues\cite{Xin-IO-CLUSTER2019, pastri} on scientific research because of the limited storage capacity and I/O bandwidth of parallel file systems in production facilities. The Hardware/Hybrid Accelerated Cosmology Code (HACC) \cite{hacc}, for example, may simulate up to 3.5 trillion particles that leads to 60 PB of data to store in one simulation; yet a system such as the Mira supercomputer [4] has only 26 PB of file system storage which is inadequate to store the simulation data. 
To make the simulation tractable, HACC researchers generally output  data in a decimation way (i.e., storing one snapshot every $K$ time steps in the simulation) which degrades the temporal consistency of simulation and loses valuable information for post-analysis. Error-controlled lossy compression techniques have been considered a better solution than the simple decimation method to reduce the data size significantly while guaranteeing the distortion of compression data is acceptable for post-analysis\cite{isc17, bigdata17, ftsz}.

There have been many data compressors (including \cite{zfp, cusz, DeepSZ, wavesz, endtoend, taoOptimizingLossyCompression2019, sz_time_based}) designed for scientific datasets. Unlike lossless compressors whose compression ratios are generally stuck with 2:1, error-bounded lossy compressors can achieve fairly high compression ratios (such as 10:1 or 100:1) while still controlling the data distortion very well\cite{sz_regression, Kai-HPDC2020, mgardx}. In order to develop or select an efficient compressor in a fair way, the compression researchers and users have to do a series of tedious work, such as collecting many state-of-the-art compressors, seeking different real-world scientific datasets and exploring sophisticated evaluation metrics.

%Lossy Compressor developers evaluate their work using various datasets on different platforms and focus on different evaluation metrics, such that it is non-trivial to understand the pros and cons of the compressors thoroughly across these works. To have a fair, comprehensive assessment of lossy compressors, researchers should run different compressors with the same group of datasets on the same platform, and analyze the result using a set of evaluation metrics commonly concerned by users. 

In this work, we propose a scientific data reduction benchmark -- SDRBench. Together with our data compression assessment tool Z-checker \cite{zchecker}, SDRBench can help compression developers and users understand the pros and cons of different compressors on various datasets. The contribution of this paper is threefold.  

\begin{itemize}
\item We collect 10+ scientific datasets across 6+ domains to support a fair, comprehensive assessment of lossy compressors. All the datasets are provided with information including description, shape, data type, and data size. In addition, some datasets are provided with physical information of the application and specific user requirements on compression errors (such as absolute error bounds and point-wise relative error bounds).
% We collected the data information together with the data providers under a close, active communication. 
\item We summarize key information related to compression techniques in the benchmark: (1) we collect the state-of-the-art lossy compressors, and analyze their pros and cons based on their design principles; (2) we analyze different types of error controls settings of the lossy compressors. (3) we summarize commonly used metrics for reduction technique assessment, including compression/decompression speed, compression ratio, peak signal to noise ratio (PSNR), rate-distortion, distribution of errors, etc.
\item Based on the data reduction benchmark and the Z-checker data assessment tool we developed, we present valuable evaluation results that can help developers and users understand the specific features of various datasets and compressors. We also provide a result analysis with six takeaways summarized. 
\end{itemize} 

The remaining of the paper is organized as follows. In Section \ref{sec:related}, we discuss the related work. In Section \ref{sec:datasets}, we describe the datasets we collected from different scientific domains and analyze their data properties. In Section \ref{sec:metrics}, we review the important metrics adopted in the benchmark. In Section \ref{sec:compressors}, we investigate the state-of-the-art compressors we collected and discuss why they are selected in our benchmark. In Section \ref{sec:evaluation}, we present our evaluation results using the benchmark and explore the data features and pros and cons of different compressors. Finally, we conclude the paper and discuss our future work in Section \ref{sec:con}.

\section{Related Work}
\label{sec:related}

Although several compression benchmarks have been developed in recent years, they mainly focus on lossless compressors, and are not suitable for assessing lossy compressors on scientific datasets. 

Squash compression benchmark\cite{squashbenchmark} and TurboBench\cite{turbobench} are two compression benchmarks that support different lossless compressors including Gzip\cite{gzip} and Zstd \cite{zstd} as plugins by constructing an abstract compression layer, making it trivial to switch between compressors. 
Large Text Compression Benchmark\cite{largetext} evaluates lossless compressors on the text data dumped from Wikipedia. It aims to encourage research on artificial intelligence and natural language processing.
The Silesia compression corpus\cite{silesia} and the Canterbury corpus\cite{canterbury} are two collections of datasets for the evaluation of lossless compressors. 

The data domains covered by the above three benchmarks and two corpora include text, source code, executable binary, PDF, image, etc. Data in those domains has different attributes from data in scientific domains (including dimension, value range, distribution, etc). Moreover, the data sizes in the lossless benchmarks are usually less than 10MB but the data sizes in scientific simulations are larger than 1GB in most cases. As a result, the above benchmarks and datasets are not suited to the evaluation of lossy compressors in scientific domains.

% Numanagic et al. \cite{nature-methods} developed a set of scripts used for measuring the performance of high-throughput sequencing (HTS) compression tools. (i.e. tools compressing FASTQ or SAM files). Since It is particularly designed for the applications of DNA sequencing and cannot be applied on the scientific simulation datasets with vast amounts of floating-point values. 
% Large Text Compression Benchmark \cite{largetext} ranks lossless data compression programs by the compressed size (including the size of the decompression program), in order to encourage research in artificial intelligence and natural language processing (NLP). This compressor cannot be applied in the lossy compression analysis for scientific datasets because it focuses only on lossless compressors and a specific problem in NLP and text compression (the ability to distinguish between high probability strings and low probability strings).
\section{The Scientific Datasets in SDRBench}
\label{sec:datasets}

\subsection{Introduction to the Scientific Datasets}

\begin{table*}[ht]
    % \vspace{-3mm}
    \centering
    \caption{Scientific Datasets}
    % \vspace{-2mm} 
    \def\arraystretch{1} 
    \footnotesize
\resizebox{2\columnwidth}{!}{        
    \begin{tabular}{|c|c|c|c|c|c|c|}
    \hline
    \textbf{Dim.} & \textbf{Name} & \textbf{Domain} & \textbf{\# files} & \textbf{Shape} & \textbf{Data Type} & \textbf{Total size}\\ \hline
    
    \multirow{2}{*}{1D} & EXAALT & Molecular dynamics simulation & 6 & 2869440 & FP32 & 60MB \\ \cline{2-7}
    &HACC & Cosmology particle simulation & 6 & 280953867 & FP32 & 5GB \\\hline
    % &NWChem & Quantum chemistry Simulation & 3 & \tiny{712996037,801098891,102953248} &FP64 & 16GB \\ \cline{2-7}
    % & Brown Samples & Synthetic & 1 & 8388609 & FP64 & 64MB \\ \hline 
    2D & CESM-ATM & Climate simulation & 100+ & 1800$\times$3600 or $26\times1800\times$3600 & FP32  & 18.5GB \\ \hline
    
    % \multirow{6}{*}{3D} & CESM-ATM (dataset2) & Climate simulation & 1 & 26$\times$1800$\times$3600 & FP32  & 17GB \\
    % \cline{2-7}
    \multirow{4}{*}{3D} & Hurricane-ISABEL & Climate simulation & 13 & 100$\times$500$\times$500 & FP32  & 1.25GB\\
    \cline{2-7}
    & NYX & Cosmology simulation & 6 & 512$\times$512$\times$512 & FP32 &3.1GB \\ \cline{2-7}
    & SCALE-LETKF &Climate simulation & 13 & 98$\times$1200$\times$1200 & FP32 & 6.4GB\\ \cline{2-7}
    & Miranda & Turbulence simulation & 7 & 256$\times$384$\times$384 &FP64 & 1GB  \\ \hline
    % & XGC & Fusion simulation & 1 & 9$\times$20694$\times$512 & FP32 & 1.2GB \\ \hline
    
    % & NSTX GPI & Fusiion Gas Puff Image & 1 & 369357 $\times$ 64 $\times$ 80 & FP64 & 4.1GB \\ \hline

    \multirow{2}{*}{4D} & EXAFEL & Images from the LCLS instrument & 1 & 10$\times$32$\times$185$\times$388 & INT16 & 51MB \\ \cline{2-7}
    & QMCPack & Quantum Structure & 1 & 288$\times$115$\times$69$\times$69&FP32 &612MB \\ \hline
    
    \end{tabular}}
    \label{tab:dataset information}
    % \vspace{-6mm}
\end{table*}

\begin{figure}[ht] \centering
\hspace{-5mm}
\subfigure[{CESM-ATM}]{
\includegraphics[width=4cm]{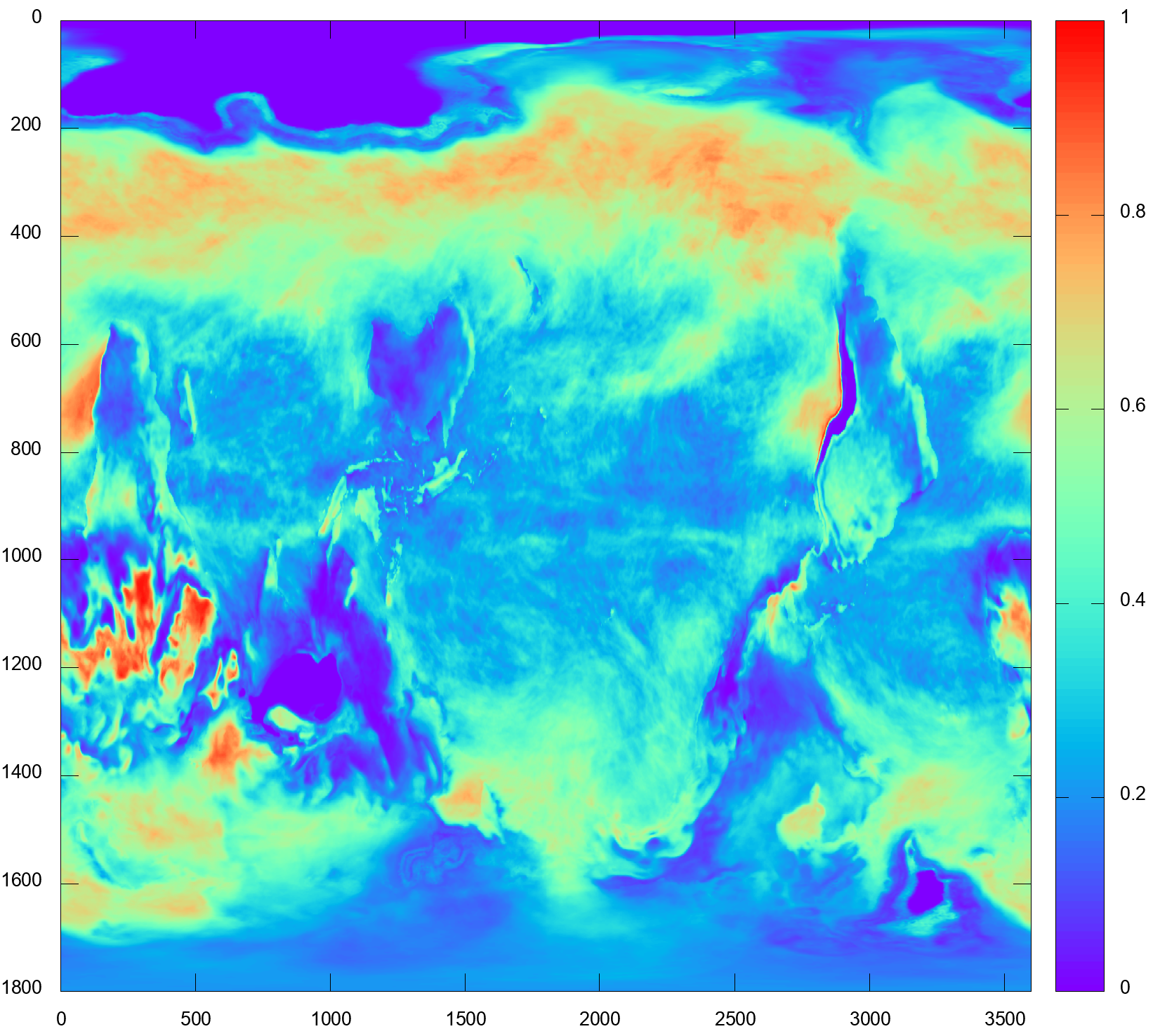}
}
\hspace{-3mm}
\subfigure[{EXAFEL}]{
\includegraphics[width=4.3cm]{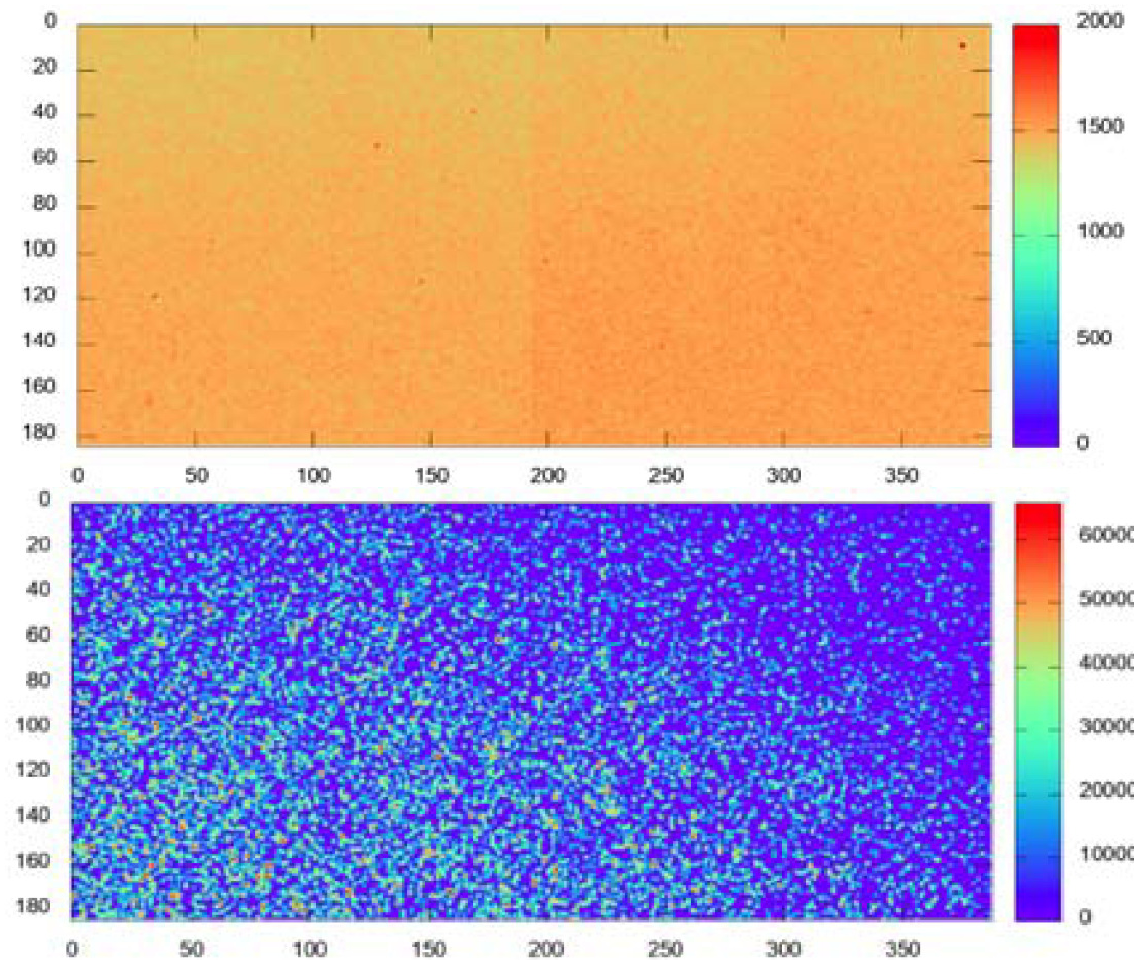}
}
\hspace{-8mm}

% \vspace{-2mm}
\caption{Visualization of CESM-ATM and EXAFEL}
\label{fig:vis}
\end{figure}

In our benchmark, we collect 10+ scientific datasets which were generated by real-world simulations. Each of the corresponding simulations/applications may potentially produce extremely large volume of data. In this section, we describe the detailed information of nine datasets in our benchmark. The summary of the nine datasets is shown in Table \ref{tab:dataset information}. 
% In this section, we describe the  in our benchmark and analyze their data properties due to the space limitation. 

\begin{figure}[ht] \centering
\hspace{-10mm}
\subfigure[{ Field: QGRAUP}]{
\raisebox{-0mm}{\includegraphics[width=4.3cm]{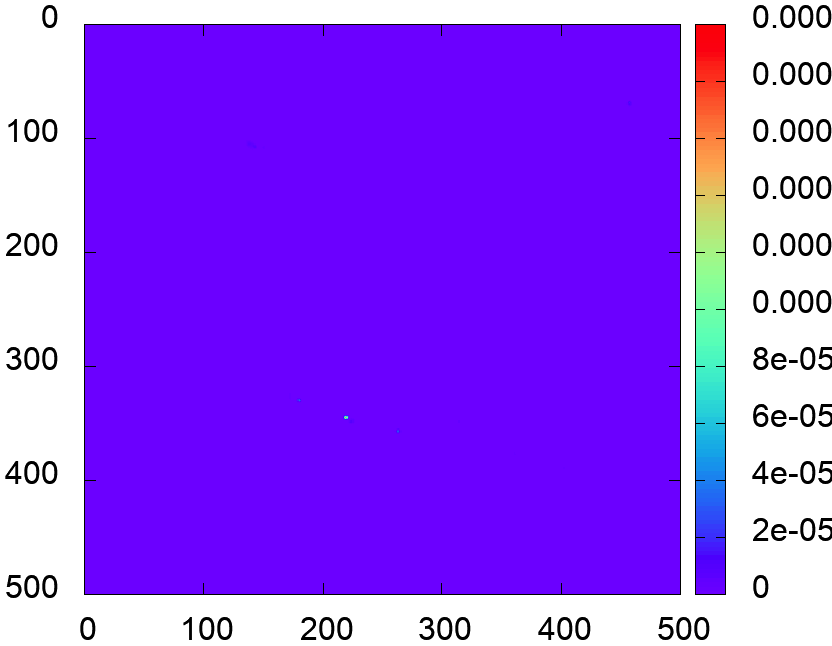}}}
\hspace{-2mm}
\raisebox{-0cm}{\subfigure[{Field: QGRAUP (log-scale)}]{
\includegraphics[width=4cm]{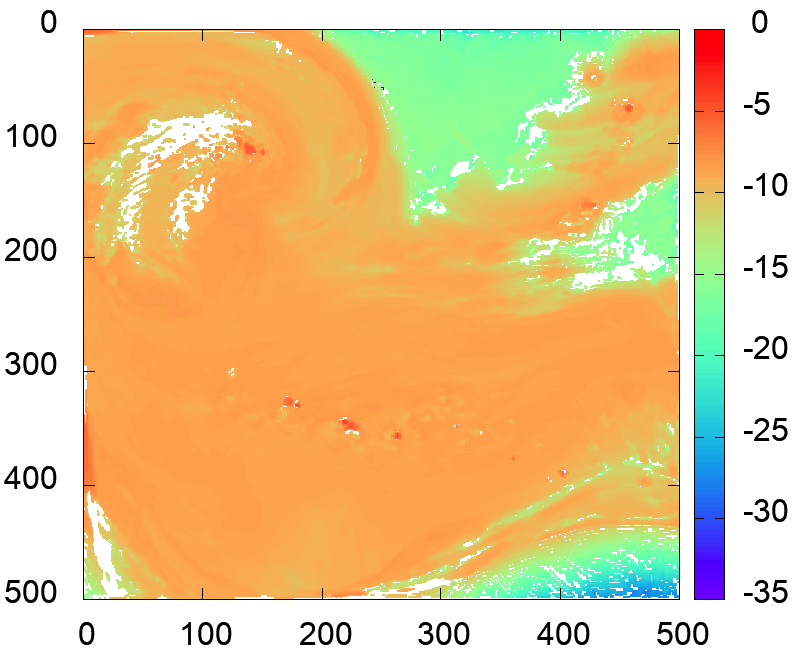}}}
\hspace{-8mm}

\hspace{-10mm}
\subfigure[{Field: U}]{
\includegraphics[width=4cm]{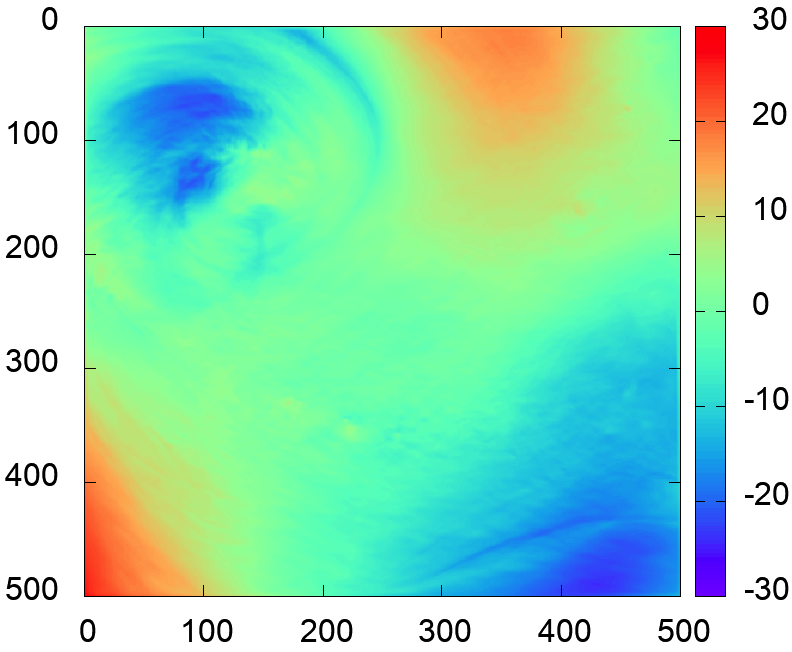}}
\hspace{0mm}
\subfigure[{Field: P}]{
\includegraphics[width=4.1cm]{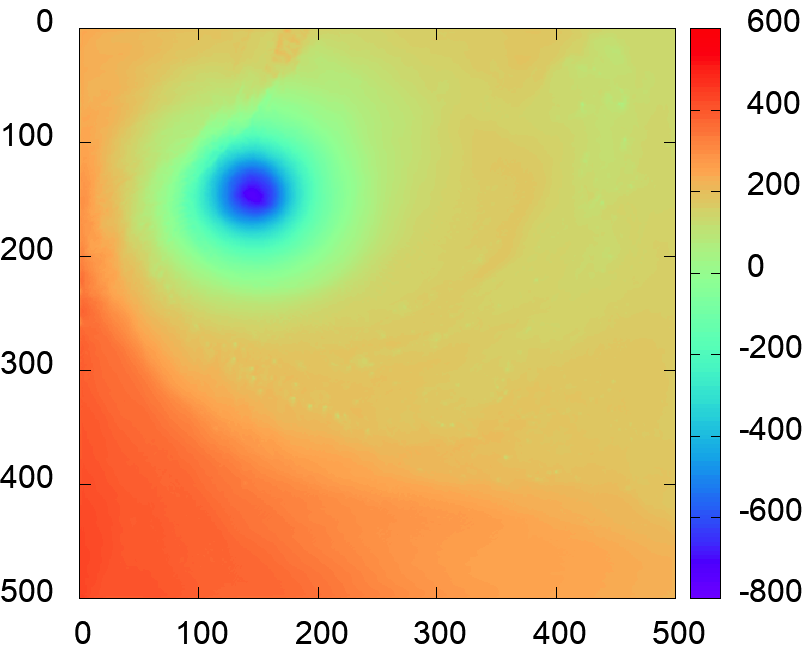}}
\hspace{-10mm}

% \vspace{-2mm}
\caption{Visualization of Hurricane-ISABEL}
\label{fig:vis-hurricane}
\end{figure}

\begin{figure}[ht] \centering
\hspace{-10mm}
\subfigure[{Field: QR}]{
\raisebox{-0mm}{\includegraphics[width=4.35cm]{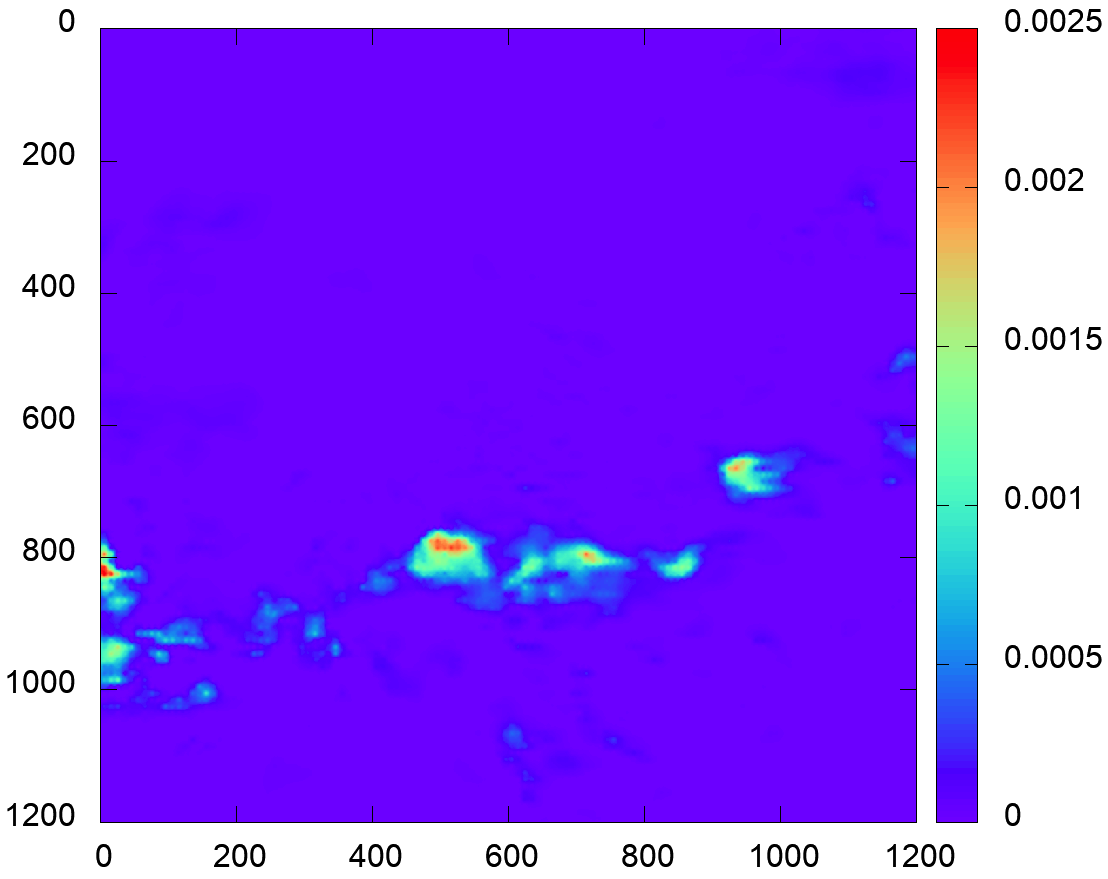}}}
\hspace{-3mm}
\raisebox{-0cm}{\subfigure[{Field: QR (log-scale)}]{
\includegraphics[width=4.05cm]{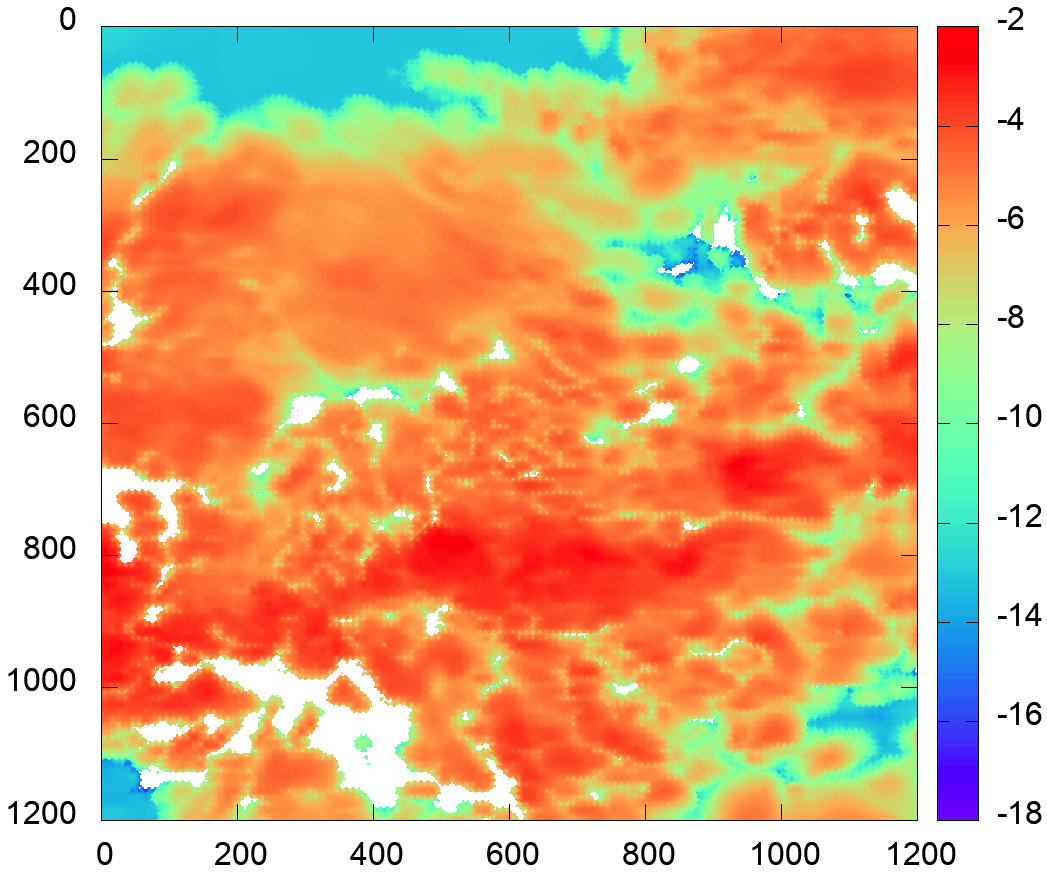}}}
\hspace{-8mm}

\hspace{-10mm}
\subfigure[{Field: U}]{
\raisebox{-1mm}{\includegraphics[width=3.95cm]{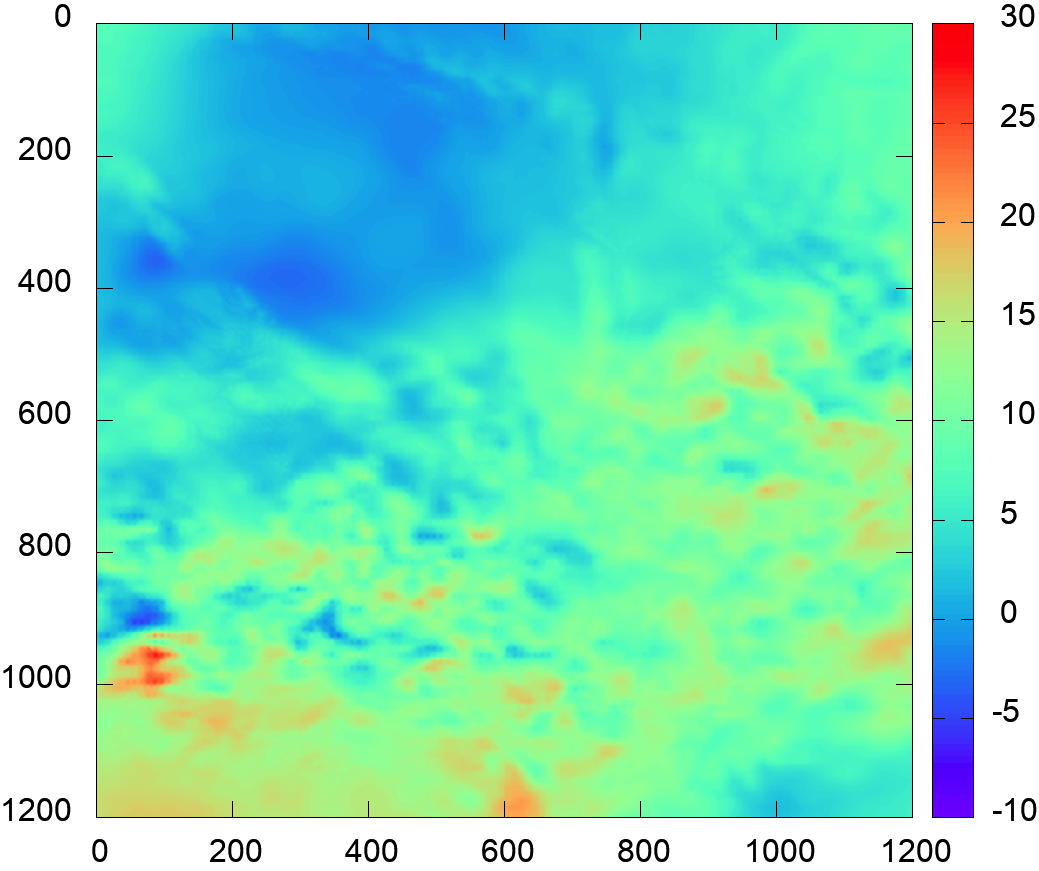}}}
\hspace{0mm}
\subfigure[{Field: PRES}]{
\raisebox{-1mm}{\includegraphics[width=4.1cm]{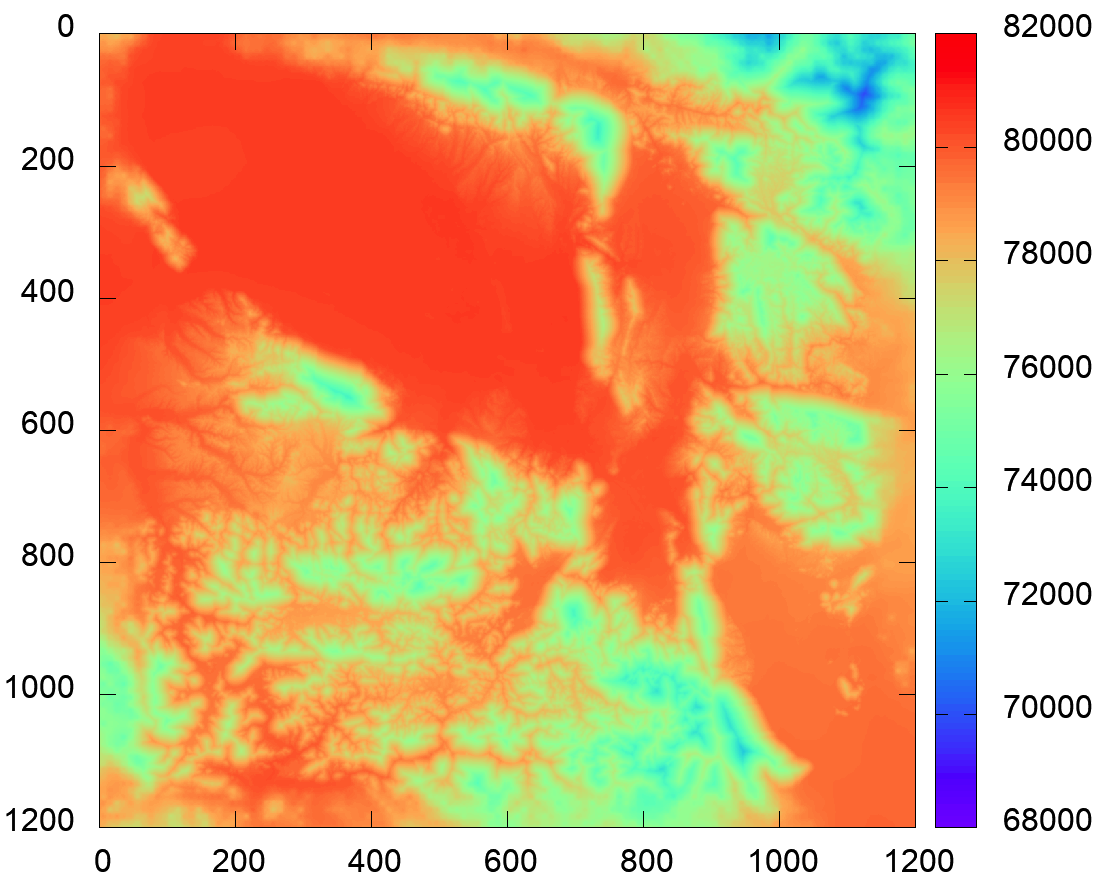}}}
\hspace{-10mm}

% \vspace{-2mm}
\caption{Visualization of SCALE-LETKF}
\label{fig:vis-scale}
\end{figure}

% \begin{itemize}
\textbf{Climate simulation} is a typical example that may produce extremely large amounts of data \cite{sz17,sz16}. We include three different climate simulation datasets in our benchmark. 
The first dataset is from the climate research project at Argonne National Laboratory. It is called CESM-ATM because it was produced based on the climate atmosphere simulation under the Community Earth System Model (CESM). 
CESM-ATM involves 60+ snapshots(timesteps), each containing 100+ fields with different dimensions. The majority of fields are 2D floating-point arrays (presented in Fig. \ref{fig:vis} (a)). Some fields involve the third dimension which represents the number of layers, while each layer is still a 2D array. For those 3D data, it is better to compress them based on 2D format instead of 3D, to be demonstrated later in detail. 
We provide one snapshot of data (out of 60 snapshots) for CESM-ATM in the SDRBench, because the dataset of 60 snapshots (1.5TB in total) is too large to download for most users and different snapshots of data exhibit similar data features. The fields and total size columns in Table \ref{tab:dataset information} are for one snapshot of CESM-ATM dataset.
The second dataset, Hurricane-ISABEL, is from IEEE Visualization 2004 contest \cite{hurricane-data}. The dataset simulates the ISABEL hurricane - the strongest hurricane in the 2003 Atlantic hurricane season. The dataset contains 13 floating-point fields in single-precision, and each field is a 3D array with the shape of $100\times500\times500$. Fig. \ref{fig:vis-hurricane} demonstrates the visualization of three fields in the Hurricane-ISABEL dataset (generated by Z-checker).
The third dataset, named SCALE-LETKF\cite{scale}, is the simulation data generated by the Local Ensemble Transform Kalman Filter (LETKF) data assimilation package with the Scalable Computing for Advanced Library and Environment - Regional Model (SCALE-RM)\cite{scale-rm}. SCALE-LETKF has 13 single-precision floating-point fields each with the shape of $98\times1200\times1200$. The visualization results are shown in Fig.
\ref{fig:vis-scale}.

\begin{figure}[ht] \centering
\hspace{-10mm}
\subfigure[{ Field: dark\_matter\_density}]{
\raisebox{0mm}{\includegraphics[width=4.05cm]{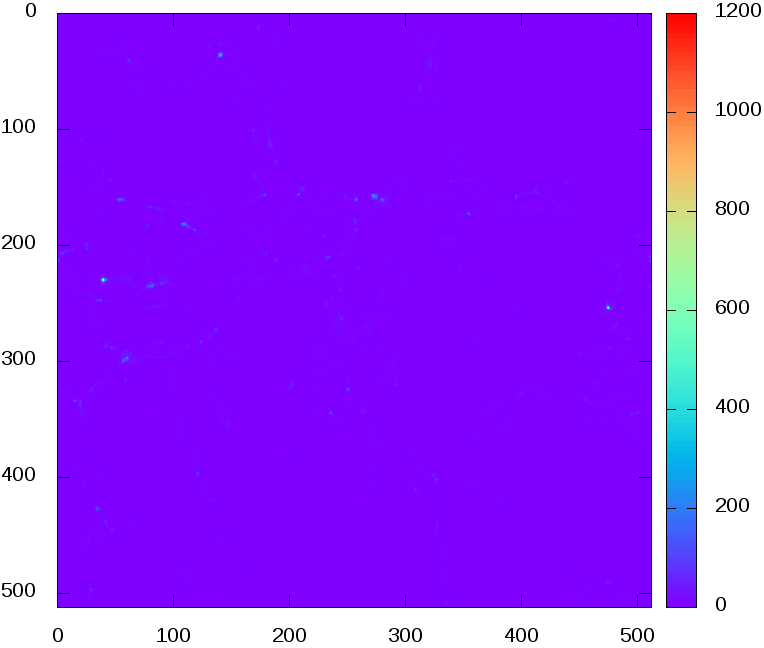}}}
\hspace{-2mm}
\raisebox{0cm}{\subfigure[{\hspace*{1mm} Field: dark\_matter\_density \newline \hspace*{15mm}(log-scale)}]{
\includegraphics[width=3.9cm]{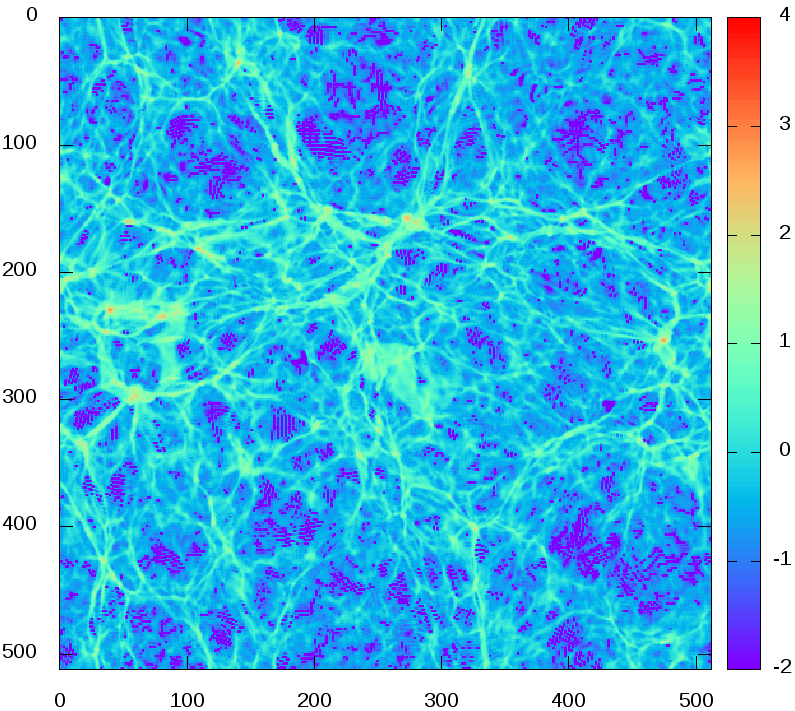}}}
\hspace{-8mm}

\hspace{-10mm}
\subfigure[{Field: temperature}]{
\raisebox{-1mm}{\includegraphics[width=4.1cm]{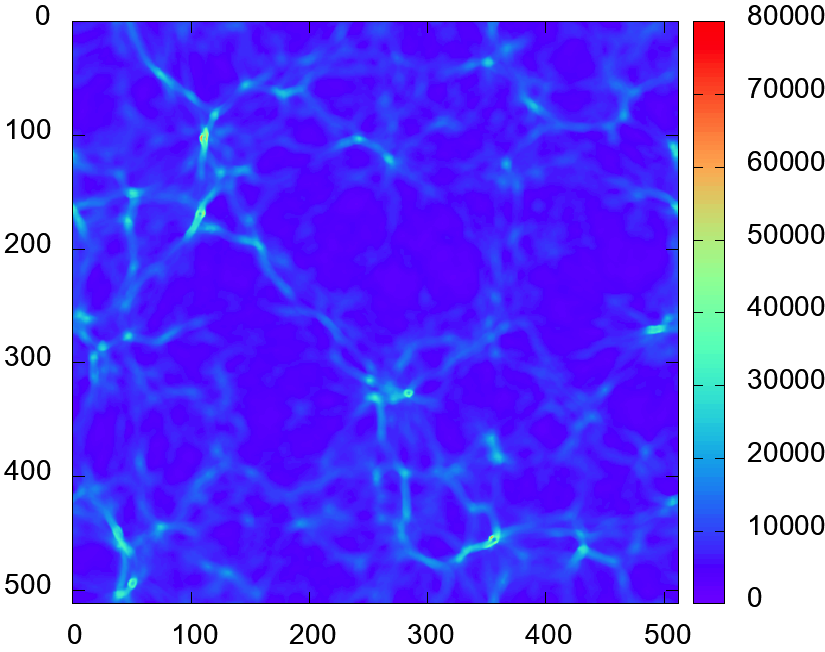}}}
\hspace{-3mm}
\subfigure[{Field: velocity\_x}]{
\raisebox{-1mm}{\includegraphics[width=4.05cm]{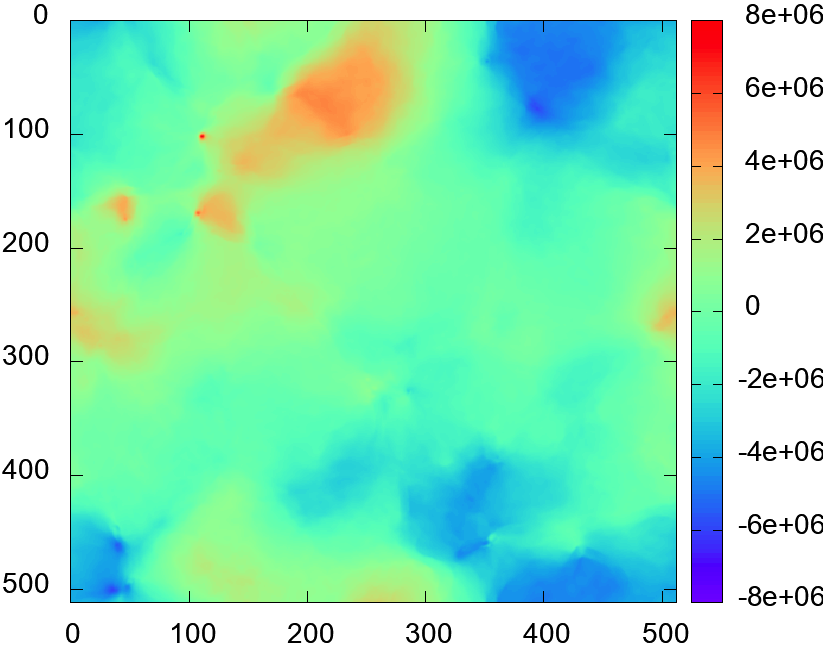}}}
\hspace{-10mm}

% \vspace{-2mm}
\caption{Visualization of NYX}
\label{fig:vis-nyx}
\end{figure}

\textbf{Cosmological N-body simulation} investigates extremely large structures such as galaxies and clusters of galaxies composed of numerous moving particles. Our benchmark involves two different cosmological simulation codes - an Hardware/Hybrid Accelerated Cosmology Code (HACC) \cite{hacc} and an adaptive mesh, compressible cosmological hydrodynamics simulation code (NYX) \cite{nyx}, both of which are widely used in the cosmological research community. The HACC data are composed of 6 1D arrays representing the position and velocity information (denoted by x, y, z, vx, vy, and vz), respectively. The NYX simulation data are post-analysis data composed of 3D arrays in space (such as dark matter density and temperature). Figure \ref{fig:vis-nyx} shows the visualization results of three fields in NYX dataset.

\begin{figure}[ht] \centering
\hspace{-8mm}
\subfigure[{Field: density}]{
\raisebox{-0mm}{\includegraphics[width=4.5cm]{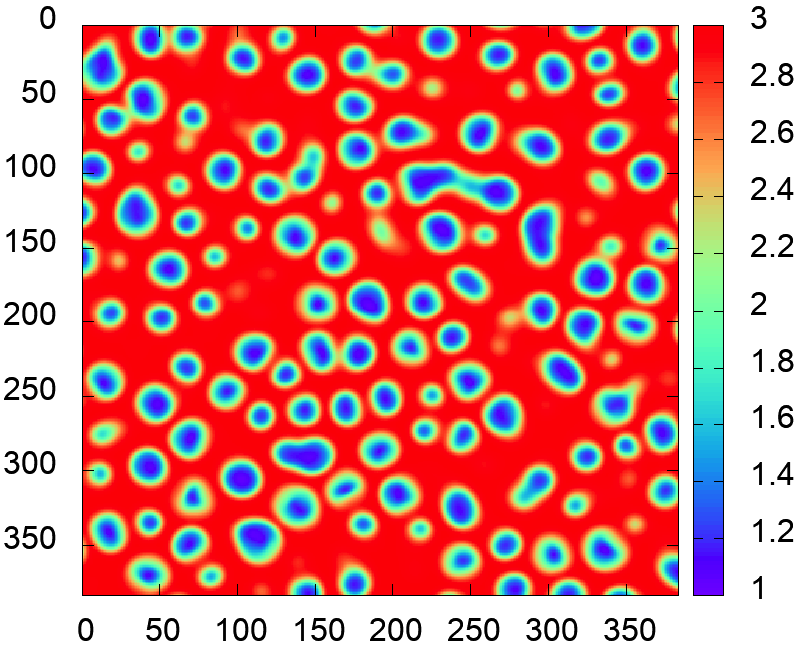}}}
\hspace{-3mm}
\raisebox{-0cm}{\subfigure[{Field: viscocity}]{
\includegraphics[width=4.5cm]{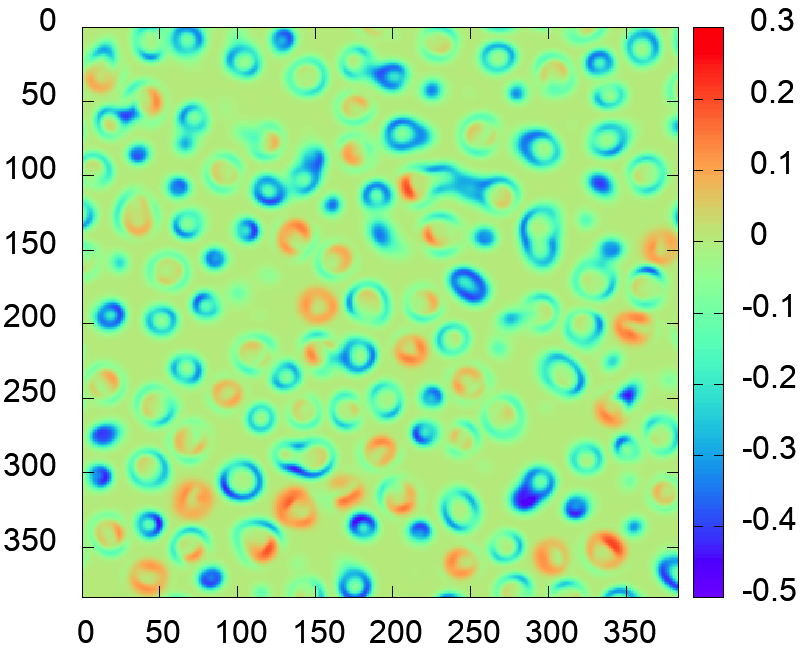}}}
\hspace{-8mm}

\hspace{-10mm}
\subfigure[{Field: velocityy}]{
\raisebox{-1mm}{\includegraphics[width=4.5cm]{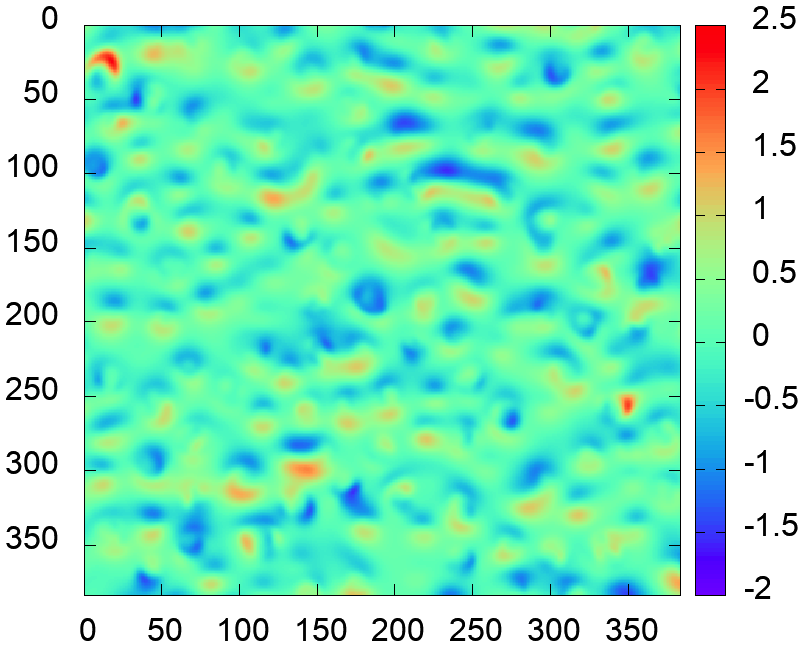}}}
\hspace{-2mm}
\subfigure[{Field: velocityz}]{
\raisebox{-1mm}{\includegraphics[width=4.35cm]{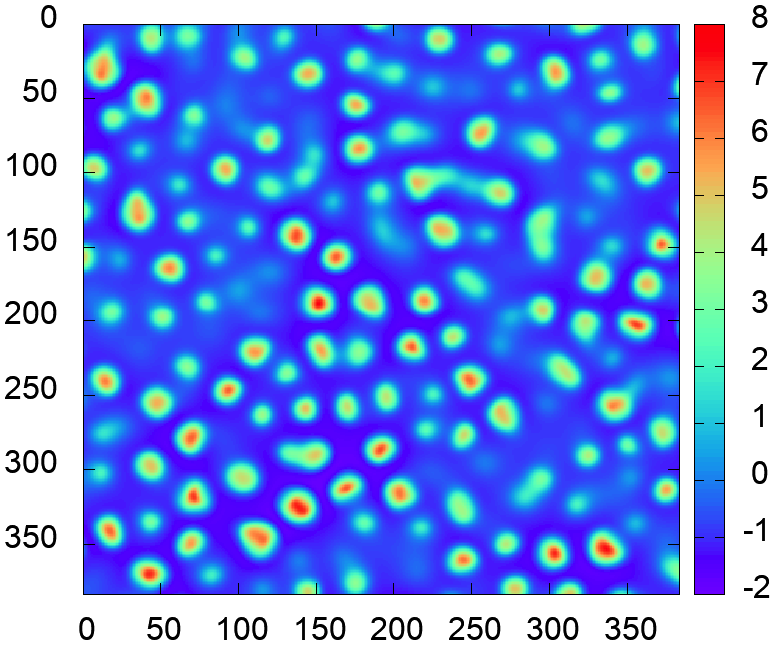}}}
\hspace{-10mm}

% \vspace{-2mm}
\caption{Visualization of Miranda}
\label{fig:vis-miranda}
\end{figure}
\textbf{Turbulence simulation} aims to solve problems regarding fluid flows by computational fluid dynamics technology. Miranda \cite{miranda}, a radiation hydrodynamics code designed for large-eddy simulation, is included in our dataset. Different from most datasets in our benchmark, the data type of Miranda is 64-bit double-precision floating-point. Miranda has seven fields, and the size of each field is $256\times 384 \times 384$. The visualization results are shown in Fig. \ref{fig:vis-miranda}.

\textbf{QMCPack} is an open source ab initio quantum Monte Carlo package for analyzing the electronic structure of atoms, molecules and solids. In the benchmark, we release the dataset (3D, single-precision) for one field called 'einspline' with two representation formats - raw data and preconditioned data, respectively. The former is the original dataset stored in the memory during the simulation and the latter reorganizes the orders of the elements because this may lead to higher compression ratios according to the data provider.  

\textbf{Molecular dynamics simulation} is also included in our benchmark, and one typical example is the Exascale Atomistic Capability for Accuracy, Length, and Time (EXAALT) project \cite{exaalt}, which aims to develop a simulation tool to address key fusion and fission energy materials challenges at the atomistic level: extending the burnup of nuclear fuel in fission reactors (dynamics of defects and fission gas clusters in $UO_2$), and developing plasma-facing components (tungsten first wall) to resist the harsh conditions of fusion reactors.

\textbf{Instrument data} is also covered by our benchmark, in addition to simulation datasets. Specifically, our benchmark provides the datasets generated/analyzed by EXAFEL project \cite{exafel}. These data was produced by Linac Coherent Light Source (LCLS) - a free-electron laser facility located at SLAC National Accelerator Laboratory. They are used to analyze biological structures in unprecedented atomic detail for modeling proteins that play a key role in many biological functions. In the benchmark, there are three fields (called 'dark', 'raw', and 'calibrated') provided by the LCLS researcher. Every field has the dimension of $10\times32\times185\times388$, which means 10 3D-images ($32\times185\times388$), where `32' is for separate panels in one detector. 
Fig. \ref{fig:vis} (b) visualizes one panel ($185\times388$) of the 'dark' field and 'calibrated' field, respectively.

\subsection{Characterization of the Scientific Datasets}
% \label{sec:characterization}

\begin{table*}[ht]
\caption{Properties of NYX Simulation Data} \centering
\vspace{-2mm}
\footnotesize
\resizebox{1.8\columnwidth}{!}{        
\begin{tabular}{|c|c|c|c|c|c|c|c|c|c|c|}
\hline
\multirow{2}{*}{\textbf{Field}} & \multirow{2}{*}{\textbf{Min}} & \multirow{2}{*}{\textbf{Avg}} & \multirow{2}{*}{\textbf{Max}} & \multirow{2}{*}{\textbf{Range}} & \multirow{2}{*}{\textbf{Entropy}} & \multicolumn{3}{c|}{\textbf{Percentile}} & \multicolumn{2}{c|}{\textbf{Autocorrelation}} \\ 
\cline{7-11} & & & & & & \textbf{1\%} & \textbf{99\%} & \textbf{range} & \textbf{lag=1} & \textbf{lag=10} \\
\hline velocity\_x & -50417k & 353.9 & 31867k & 82283k & 7.36 & -14932k & 14032k & 28965k & 1.000 & 0.991 \\
\hline velocity\_y & -43933k & 52.97 & 56506k & 100438k & 7.39 & -11700k & 12729k & 24429k & 1.000 & 0.984 \\
\hline velocity\_z & -38938k & -65.7 & 33386k & 72324k & 7.37 & -9340k & 7863k & 17204k & 0.998 & 0.919 \\
\hline temperature & 2281 & 8453.3 & 4783k & 4780k & 7.01  & 3610 & 27984 & 24373 & 0.916 & 0.331 \\
\hline dark\_matter\_density & 0 & 1 & 13779 & 13779 & 7.19 & 0.000 & 10.457 & 10.457 & 0.775 & 0.038\\
\hline baryon\_density & 0.0580 & 1 & 115863 & 115862 & 7.1 & 0.137 & 7.420 & 7.283 & 0.568 & 0.003 \\
\hline 
\end{tabular}
}
\label{tab:propAnalysis-nyx}
\end{table*}

In this section, we characterize the scientific datasets in the benchmark. Due to the space limitation, we just present the results of some fields in NYX, Hurricane-ISABEL, and SCALE-LETKF, and the full reports can be found on the Z-checker website \cite{zchecker}.
%, Moledular dynamics simulation data (EXAALT), and LCLS instrument data (EXAFEL). 

\textbf{NYX.}  Table \ref{tab:propAnalysis-nyx} presents the data properties of NYX simulation. We can observe that among the six fields, three of them have more than 5000X larger value range than the rest fields. As a result, the value range has to be taken into account if an absolute error bound is used in the compression, otherwise either the compression ratio would be too small or the compression errors would be too large to be available for the post-analysis. 

Some datasets such as the dark matter density of NYX should be transformed to log-scale data before performing the visualization analysis, as confirmed by the data providers. Fig. \ref{fig:vis-nyx} shows one slice image of the dark matter density data based on the original scale and log-scale, respectively. The log-scale data is much easier for the visualization than the original scale, because this dataset spans a very large value range ([0,13779]) while the majority of data are pretty small (the range between 1\% percentile and 99\% percentile is only 10.45). Accordingly, we adopt the compression based on the log-scale data for such fields (including dark matter density and baryon density), and other fields still apply the original-scale compression.

The autocorrelation analysis indicates that the velocity fields have higher correlation than other fields. The lag 10 autocorrelation of baryon\_density is only 0.003 while the lag 10 autocorrelation of velocity\_x is 0.991. Therefore, using the same lossy compressor, baryon\_density may exhibit a different compression quality compared to velocity\_x .

\begin{table*}[ht]
\caption{Properties of Hurricane-ISABEL Simulation Data} \centering
\vspace{-2mm}
\footnotesize
\resizebox{1.8\columnwidth}{!}{        
\begin{tabular}{|c|c|c|c|c|c|c|c|c|c|c|}
\hline
\multirow{2}{*}{\textbf{Field}} & \multirow{2}{*}{\textbf{Min}} & \multirow{2}{*}{\textbf{Avg}} & \multirow{2}{*}{\textbf{Max}} & \multirow{2}{*}{\textbf{Range}} & \multirow{2}{*}{\textbf{Entropy}} & \multicolumn{3}{c|}{\textbf{Percentile}} & \multicolumn{2}{c|}{\textbf{Autocorrelation}} \\ 
\cline{7-11} & & & & & & \textbf{1\%} & \textbf{99\%} & \textbf{range} & \textbf{lag=1} & \textbf{lag=10} \\
\hline W & -3.241 & 0.0038 & 13.3332 & 16.574 & 4.489 & -0.278 & 0.456 & 0.734 & 0.703 & 0.148 \\
\hline V & -45.615 & 3.5531 & 48.0858 & 93.7 & 8.914  & -22.666 & 32.132 & 54.798 & 0.998 & 0.978\\
\hline U & -53.023 & -2.223 & 39.56 & 92.581 & 8.638 & -29.244 & 20.281 & 49.525 & 0.996 & 0.950 \\
\hline P & -3411.741 & 375.94 & 3224.4 & 6636.1 & 7.747 & -738 & 2018 & 2756 & 0.998 & 0.984 \\
\hline TC & -76.554 & -30.793 & 29.647 & 106.201 & 9.682  & -72.767 & 25.076 & 97.843 & 1.000 & 1.000 \\
\hline CLOUD & 0 & 8.6E-06 & 0.00205 & 0.002 & 1.0222 & 0 & 0.000253 & 0.000253 & 0.809 & 0.396 \\
\hline PRECIP & 0 & 1.24E-05 & 0.00751 & 0.008 & 0.891 & 0 & 0.000314 & 0.000314 & 0.864 & 0.428 \\
\hline QCLOUD & 0 & 6.4E-06 & 0.00205 & 0.002 & 0.484  & 0 & 0.000235 & 0.000235 & 0.803 & 0.393 \\
\hline QVAPOR & 0 & 0.0023 & 0.02 & 0.02 & 6.397  & 0 & 0.0168 & 0.0168 & 0.998 & 0.984\\
\hline QGRAUP & 0 & 3.8E-06 & 0.0073 & 0.0073 & 0.38 & 0 & 0.000115 & 0.000115 & 0.857 & 0.380 \\
\hline
\end{tabular}
}
\label{tab:propAnalysis-hurricane}
\end{table*}

\textbf{Hurricane-ISABEL.} Table \ref{tab:propAnalysis-hurricane} presents the data properties of the 10 fields in Hurricane ISABEL simulation. The value range differs largely on different fields. Specifically, the fields CLOUD, PRECIP, QCLOUD, QGRAUP need to be log-transformed before compression, because the original-scale data cannot be visualized directly (shown in Fig. \ref{fig:vis-hurricane} (a)).

\begin{table*}[ht]
\caption{Properties of Scale-LETKF Simulation Data} \centering
\vspace{-2mm}
\footnotesize
\resizebox{1.8\columnwidth}{!}{        
\begin{tabular}{|c|c|c|c|c|c|c|c|c|c|c|}
\hline
\multirow{2}{*}{\textbf{Field}} & \multirow{2}{*}{\textbf{Min}} & \multirow{2}{*}{\textbf{Avg}} & \multirow{2}{*}{\textbf{Max}} & \multirow{2}{*}{\textbf{Range}} & \multirow{2}{*}{\textbf{Entropy}} & \multicolumn{3}{c|}{\textbf{Percentile}} & \multicolumn{2}{c|}{\textbf{Autocorrelation}} \\ 
\cline{7-11} & & & & & & \textbf{1\%} & \textbf{99\%} & \textbf{range} & \textbf{lag=1} & \textbf{lag=10} \\
\hline U &   -71.75 & 5.79 & 46.67 & 118.43 & 7.23 & -23.9 & 18.5 & 42.4 & 0.999 & 0.992 \\
\hline QV  & 0 & 0.0043 & 0.0195 & 0.0195 & 7.392 & 1.21E-06 & 0.016 & 0.016 & 1.000 & 0.999 \\
\hline QC  & 0 & 6.37E-06 & 0.0030 & 0.0030 & 3.924 & 0 & 0.000163 & 0.000163 & 0.995 & 0.794 \\
\hline QG  & 0 & 1.46E-05 & 0.0148 & 0.0148 & 7.479 & 0 & 0.0002 & 0.0002 & 0.989 & 0.822 \\
\hline QI  & 0 & 4.17E-06 & 0.0016 & 0.0016 & 4.008  & 0 & 7.45E-05 & 7.45E-05 & 0.996 & 0.900\\
\hline PRES  & 2285 & 42841 & 101820 & 99534 & 7.106  & 2480 & 99414 & 96933 & 1.000 & 1.000\\
\hline RH &  0 & 44.64 & 204.47 & 204.47 & 7.115  & 0.151 & 97.4 & 97.2 & 1.000 & 0.991\\
\hline T &   181.91 & 251.06 & 314.63 & 132.71 & 6.621  & 202 & 305 & 104 & 1.000 & 1.000\\
\hline QR &  0 & 1.28E-05 & 0.00637 & 0.00637 & 7.500  & 0 & 0.000324 & 0.000324 & 0.995 & 0.887\\
\hline W &   -37.12 & -0.0194 & 26.77 & 63.89 & 7.448 & -4.94 & 3.58 & 8.52 & 0.997 & 0.847 \\
\hline QS &  0 & 6.43E-06 & 0.000756 & 0.000756 & 7.494  & 0 & 0.000084 & 0.000084 & 0.998 & 0.944\\
\hline V &   -40.15 & -0.412 & 59.42 & 99.57 & 7.383 & -18.6 & 23.6 & 42.2 & 0.998 & 0.972 \\
\hline
\end{tabular}
}
\label{tab:propAnalysis-scale}
\end{table*}

\textbf{SCALE-LETKF}
Table \ref{tab:propAnalysis-scale} presents the data properties of SCALE-LETKF simulation. Similar to NYX dataset, some fields including QC, QG, and QR in SCALE-LETKF dataset need to be log-transformed before compression. Fig.\ref{fig:vis-scale} (a,b) shows one slice image of the QR data based on the original scale and log-scale, respectively, and it confirms that the log-scale is better for visualization than the original scale.  
\section{The Lossy Compressor Assessment Metrics in SDRBench}
\label{sec:metrics}
In this section, we discuss the metrics to evaluate lossy compressors on scientific datasets. 

The compression developers and users generally focus on the following three metrics regarding the distortion of data -- maximum absolute error, maximum point-wise relative error, peak-to-signal noise ratio (PSNR). Maximum absolute error is defined as the maximum difference between original data and decompressed data. Maximum point-wise relative error refers to the maximum ratio of the absolute point-wise error to the original data value. PSNR is used to assess the average compression error, and its definition is shown in Formula (\ref{psnr}).	
\begin{equation}
\label{psnr}
PSNR = 20\cdot \log_{10}{(value\_range)} - 10\cdot \log_{10}{(MSE)}. 
\end{equation}
where value\_range and \emph{MSE} are referred to as the value range of the dataset and mean squared error between the original data and decompressed data. 

In addition to the above common metrics, there are some other metrics designed for specific purposes. Autocorrelation of compression errors, for instance, is used to assess the correlation of the compression errors. Distribution of compression errors is also studied to understand the overall distortion of the data in a statistical way. The structural similarity (SSIM) \cite{Wang_2004} is an index to measure the similarity between the original dataset and decompressed one. The SSIM is the product of three terms (luminance, contrast and structure) evaluating respectively the matching of intensity between the two datasets $a$ and $b$, the variability and the co-variability of the two signals. In statistical terms, luminance, contrast, and structure can be seen as evaluating the bias, variance, and correlation between the two datasets, respectively. SSIM is expressed as
$$S\hspace{-0.2mm}S\hspace{-0.2mm}I\hspace{-0.2mm}M(a,b) \hspace{-1mm}=\hspace{-1mm} \underbrace{\left(\frac{2\mu_{a}\mu_{b} + c_1}{\mu_{a}^{2}\hspace{-0.4mm} +\hspace{-0.4mm} \mu_{b}^{2} \hspace{-0.4mm}+\hspace{-0.4mm} c_{1}} \right)}_{\rm luminance} \underbrace{\left( \frac{2\sigma_{a}\sigma_{b} + c_2}{\sigma_{a}^{2} \hspace{-0.4mm}+\hspace{-0.4mm} \sigma_{b}^{2} \hspace{-0.4mm}+\hspace{-0.4mm} c_{2}}  \right)}_{\rm contrast} \underbrace{\left(\frac{\sigma_{ab}\hspace{-0.4mm} +\hspace{-0.4mm} c_3}{\sigma_{a}\sigma_{b} \hspace{-0.4mm}+\hspace{-0.4mm} c_{3}}   \right)}_{\rm structure},$$
with $\mu_{x}$, $\sigma_{x}$ and $\sigma_{xy}$, respectively, being the mean, standard deviation and the cross-covariance of each dataset, and $c_1$, $c_2$ and $c_3$ are constants derived from the datasets. 
SSIM takes values between $-1$ and $1$, and the closer to $1$, the more similar the two signals are. In Z-checker, we provide two versions of SSIM, one for 1D dataset (directly using the above formula) and the other for 2D dataset (calculating mean SSIM based on every data point in the dataset \cite{vis-ssim}), respectively. 
%It is common as well to investigate the three components (luminance, contrast, variability) separately. 

%The Kolmogorov–Smirnov test (KS-test) is a statistical test to equality the between two probability distributions. It is based on the Kolmogorov-Smirmov distance that quantifies the distance between two cumulative distribution functions (cdf), which are in our case the empirical cdf of the original dataset and the empirical cdf of decompressed data. 

% \section{Z-checker: A Lossy Data Compression Assessment Tool}
% \label{zchecker}

All the evaluation metrics above are included in our compression assessment tool Z-checker.
Z-checker helps lossy compressor developers and users explore the features of scientific datasets and understand the data alteration after compression in a systematic and reliable way. On the one hand, Z-checker combines a group of data analysis components for data compression. On the other hand, Z-checker is implemented as an open-source community tool to which users and developers can contribute and add new analysis components based on their needs. 
% We have integrated all the evaluation metrics in Z-chcecker. Specifically, 
For lossy compressor developers, Z-checker can be used to characterize critical properties (such as entropy, distribution, power spectrum, principal component analysis, and autocorrelation) of any dataset. For lossy compression users, Z-checker can analyze the compression quality (PSNR, normalized MSE, rate-distortion, rate-compression error, spectral, distribution, derivatives) and provide statistical analysis of the compression errors (maximum, minimum, and average error, autocorrelation, distribution of errors). Z-checker can also be extended with more plugins coded in other programming languages/libraries, such as R and FFTW3.

\section{The Lossy Compressors in SDRBench}
\label{sec:compressors}

Since scientific applications often have strict requirements on the distortion of compression data, our benchmark mainly focuses on the error-controlled lossy compressors.
%, while other non-error-controlled compression techniques (such as downsampling and interpolation) are also integrated into our framework for the purpose of comparison. 
In the following, we describe three state-of-the-art lossy compressors due to the space limitation, and more compressors can be found on the website of our benchmark.

\begin{itemize}
\item \textbf{SZ} \cite{sz17,sz16} is an error-bounded compressor, which contains four critical steps: (1) predict the value of each data point; (2) perform linear-scaling quantization; (3) perform customized variable-length encoding; and (4) perform optional lossless compression by compressors such as Zstd \cite{zstd}. Its particular advantage is allowing users to customize their own data prediction methods based on the data features such that the compression quality could be improved significantly for specific datasets. SZ provides three ways to control the compression errors, including absolute error bound, point-wise relative error bound, and peak-to-signal noise ratio (PSNR). In our assessment, we adopt the latest version which is SZ 2.1.
\item \textbf{ZFP} \cite{zfp} is another error-bounded lossy compressor supporting random access during the decompression because of its block-wise design. It contains five critical steps: (1) align the values in each block to a common exponent; (2) convert the floating-point values to a fixed-point representation; (3) decorrelate values by applying orthogonal transforms; (4) order the transform coefficients; and (5) perform an embedded coding algorithm. ZFP allows users to set an absolute error bound for the compression or specify an integer number (called \emph{precision}) to obtain a point-wise relative error bounding effect.
\item \textbf{SZ(Hybrid)} \cite{xin-sc19-hybrid-sz-zfp} is a hybrid lossy compressor that integrates a transform-based predictor into the SZ compressor framework. Compared with ZFP which also utilizes the transform-based technology, SZ(Hybrid) has better compression quality on high compression ratio cases because it optimizes the encoding strategy of the transform-based predictor. SZ(Hybrid) adopts a rate-distortion estimation process on the sampling data to select the best predictor between the transform-based predictor and data-fitting-based predictor. The estimation process incurs 50\%$\sim$ 100\% runtime overhead compared with SZ.    
% \item \textbf{FPZIP} \cite{fpzip} also adopts the prediction+quantization model. It adopts a different quantization approach and encoding method, which makes it more suitable for point-wise relative error bounded compression, unlike SZ which was designed for absolute error bound originally. Hence, FPZIP provides only the precision bounding mode.
\end{itemize}   

\section{Evaluation of Lossy Compressors}
\label{sec:evaluation}

In this section, we analyze the three lossy compressors and summarize six takeaways. We focus on critical metrics based on six datasets (NYX, QMCPack, Hurricane-ISABEL, CESM-ATM, EXAFEL, and SCALE-LETKF) due to the space limitation. 
% EXAFEL data were 16-bit integer-values originally and it is converted to floating-point values in our evaluation. 
All compression results were generated by running SZ v2.1, ZFP v0.5.5 and SZ(Hybrid)\footnote{ Available at https://github.com/lxAltria/hybrid\_lossy\_compression} on Bebop server \cite{bebop}. 
% We will show that SZ and ZFP are the two best compressors in class, and their compression quality depends on datasets and none of them is always better than the other. 

\begin{figure}[ht] \centering
\hspace{-6mm}
\subfigure[{REB=1E-2}]{
\includegraphics[width=4.6cm]{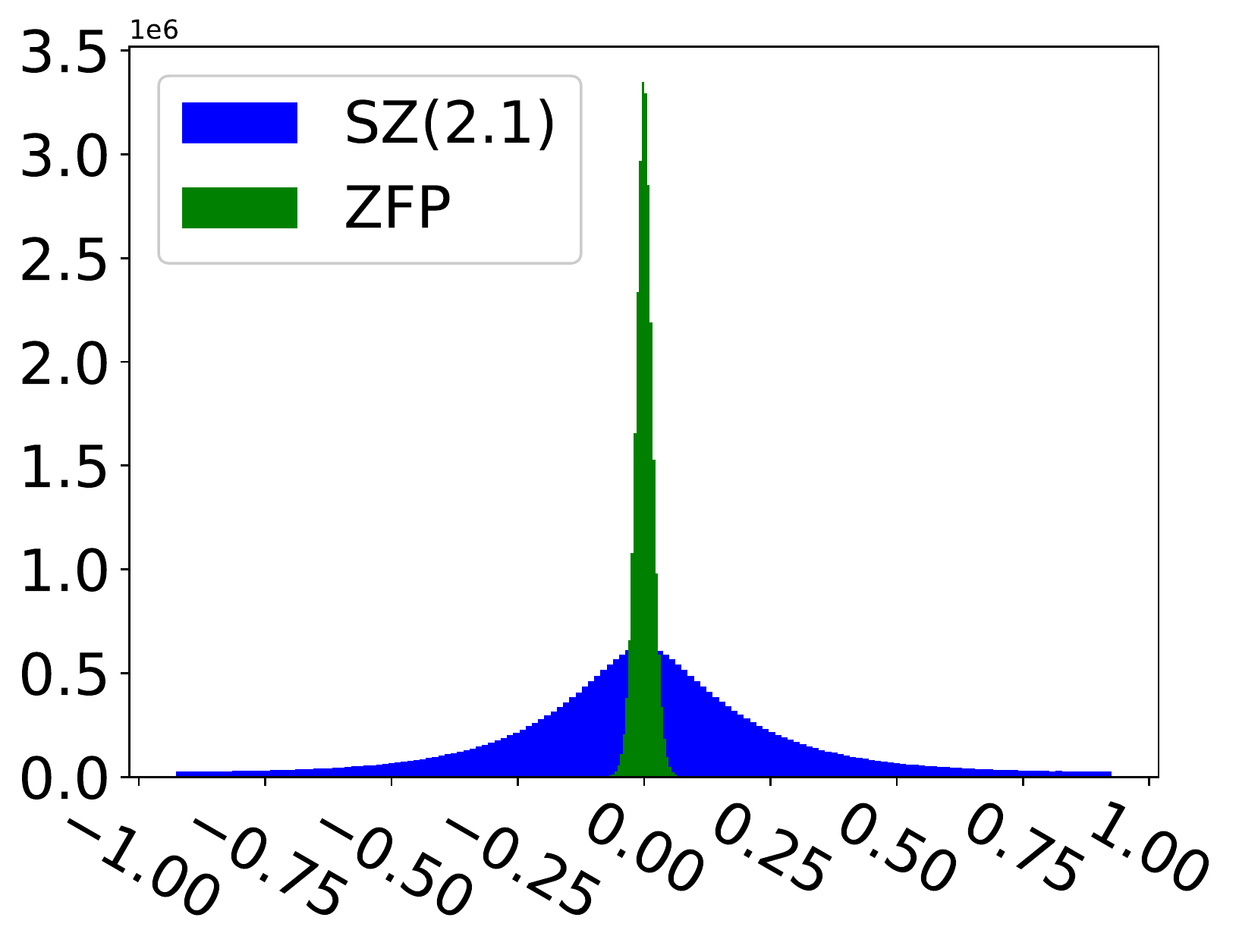}}
\hspace{-6mm}
\subfigure[{REB=1E-3}]{
\raisebox{-1.2mm}{\includegraphics[width=4.6cm]{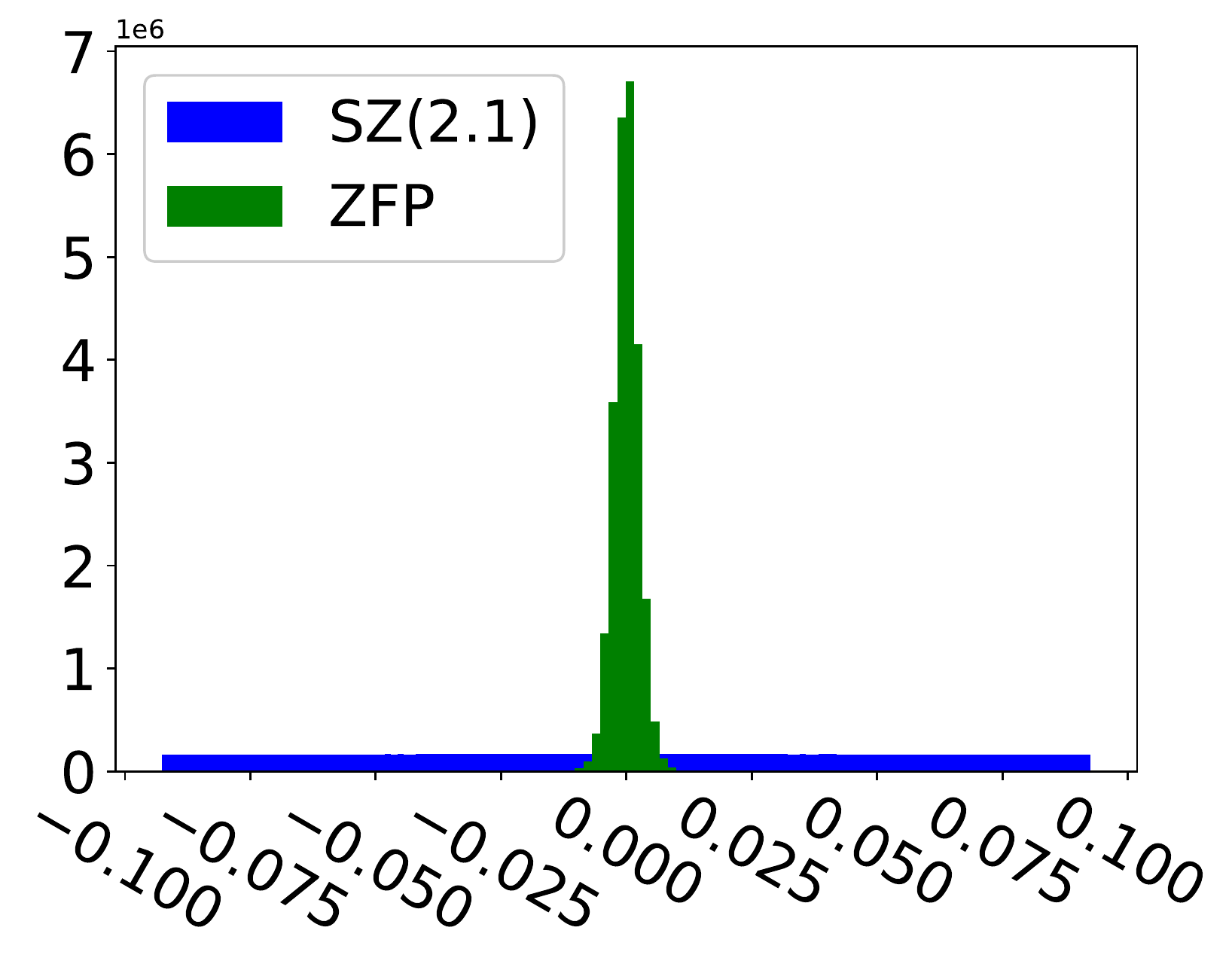}
}}
\hspace{-8mm}

% \vspace{-2mm}
\caption{Error Distribution (Hurricane (U))}
\label{fig:error-hurricane}
\end{figure}
We first verify that the compression errors are within the user defined error bound for all the lossy compressors. Fig. \ref{fig:error-hurricane} shows the error distribution of field U in Hurricane-ISABEL dataset. The value range of field U is 92.58. The absolute error bounds of U are 0.92 and 0.09 for value-range-based relative error bounds\footnote{Value-range-based relative error bound is defined as the ratio of absolute error bound to the data value range.} 1E-2 and 1E-3, respectively. The figure confirms that the compression errors are within the absolute error bound. \textbf{Takeaway 1: Compression Error.} The compression errors of SZ have different distributions with different error bound settings. ZFP tends to over preserve the compression precision so that the maximum compression error is much smaller than the error bound.

\begin{figure*}[ht] \centering
%\hspace{-6mm}
\subfigure[{REB = 1E-2}]{
\raisebox{-2mm}{\includegraphics[width=5cm]{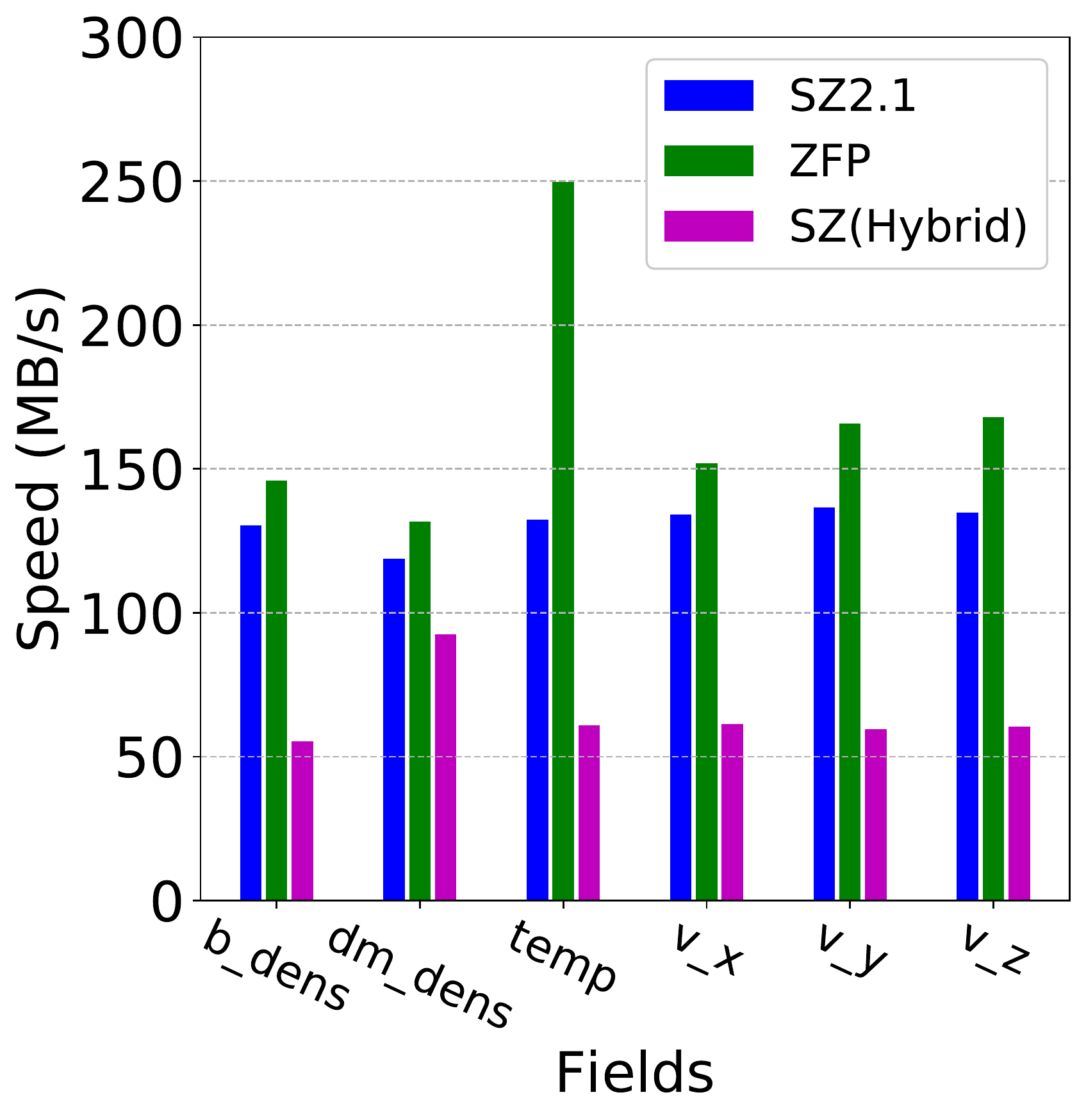}
}}
%\hspace{-4mm}
\subfigure[{REB = 1E-4}]{
\raisebox{-2mm}{\includegraphics[width=5cm]{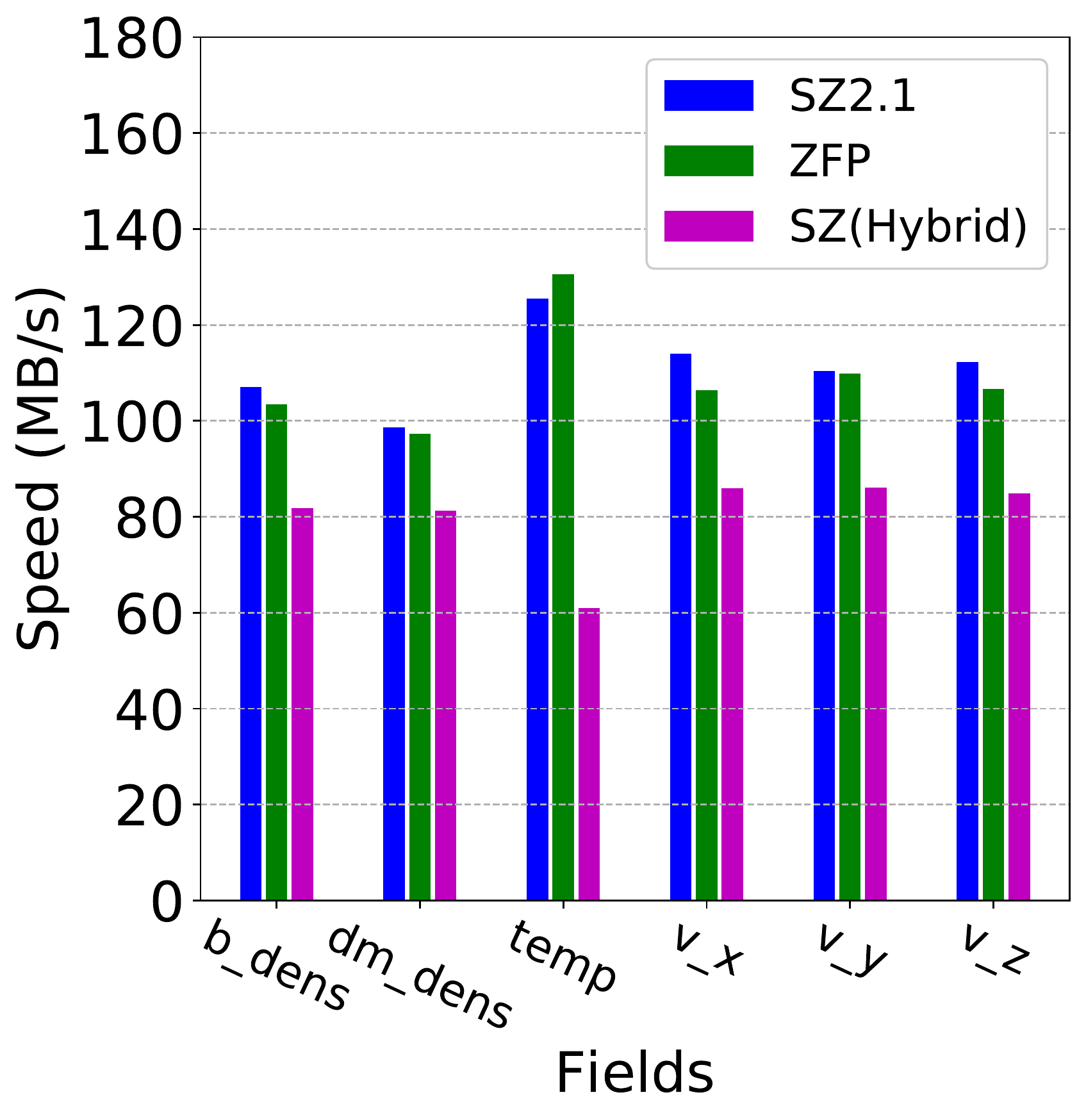}
}}
%\hspace{-2mm}

%\vspace{-2mm}
\caption{Compression Speed (NYX)}
\label{fig:crate}
\end{figure*}

\begin{figure*}[ht] \centering
%\hspace{-6mm}
\subfigure[{REB = 1E-2}]{
\raisebox{-2mm}{\includegraphics[width=5cm]{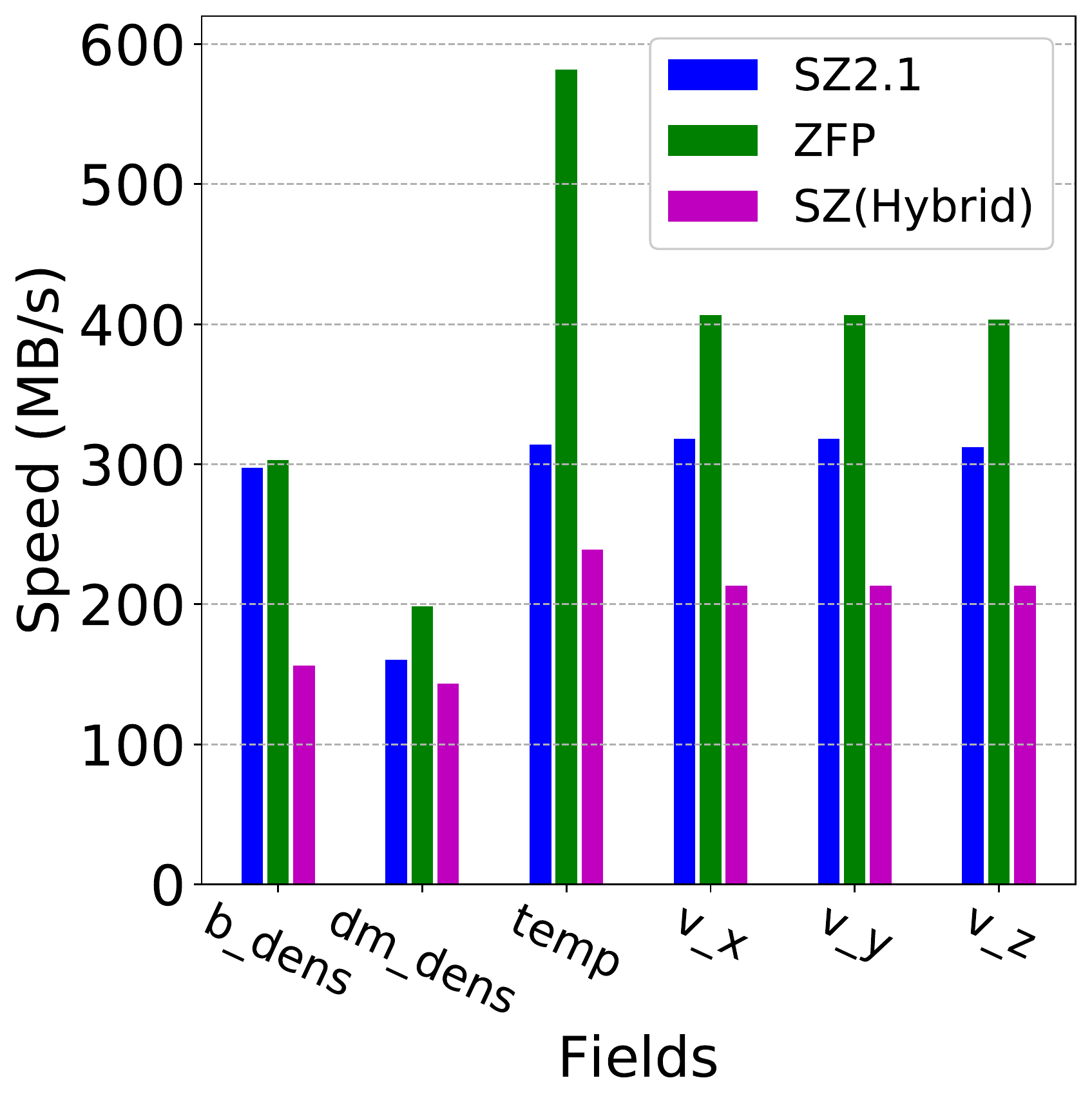}}
}
%\hspace{-4mm}
\subfigure[{REB = 1E-4}]{
\raisebox{-2mm}{\includegraphics[width=5cm]{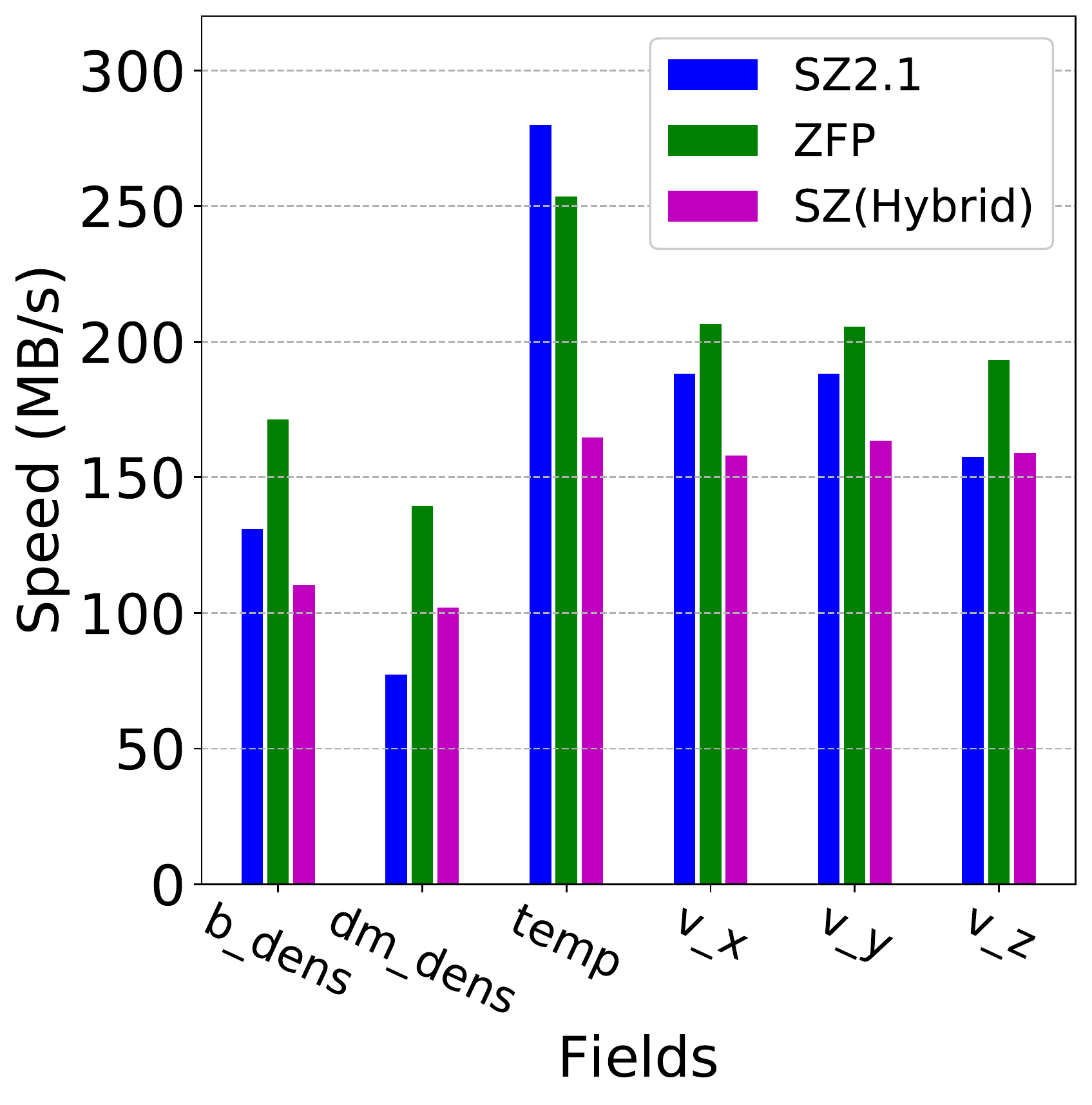}
}}
%\hspace{-2mm}

%\vspace{-2mm}
\caption{Decompression Speed (NYX)}
\label{fig:drate}
\end{figure*}

Fig. \ref{fig:crate} and Fig. \ref{fig:drate} present the compression/decompression speed under the value-range-based relative error bound of 1E-2 and 1E-4, respectively. 
% Since FPZIP controls errors by a precision setting instead of absolute error, we ran it with different precision values (12$\sim$25) and select the results with the corresponding maximum absolute errors for the comparison with SZ and ZFP. 
\textbf{Takeaway 2: Compression Speed.} It is observed that ZFP is about $10\% \sim 100\%$ faster than SZ , and SZ is about $10\% \sim260\%$ faster than SZ(Hybrid). 

\begin{figure*}[ht] \centering
% \resizebox{1.8\columnwidth}{!}{        

\hspace{-6mm}
\subfigure[{CLOUD}]{
\raisebox{-2mm}{\includegraphics[width=4.6cm]{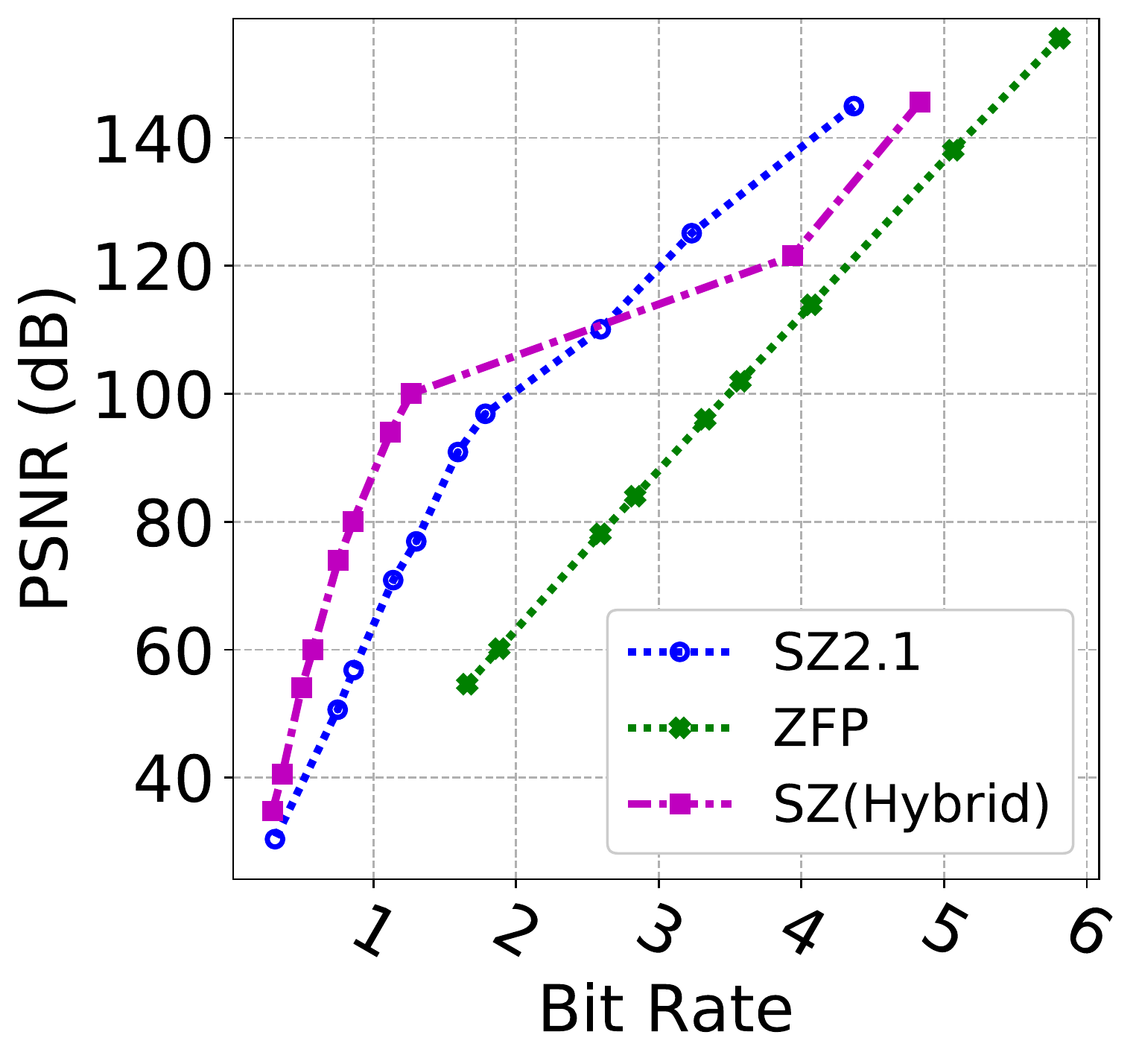}
}}
\hspace{-6mm}
\subfigure[{PRECIP}]{
\raisebox{-2.6mm}{\includegraphics[width=4.5cm]{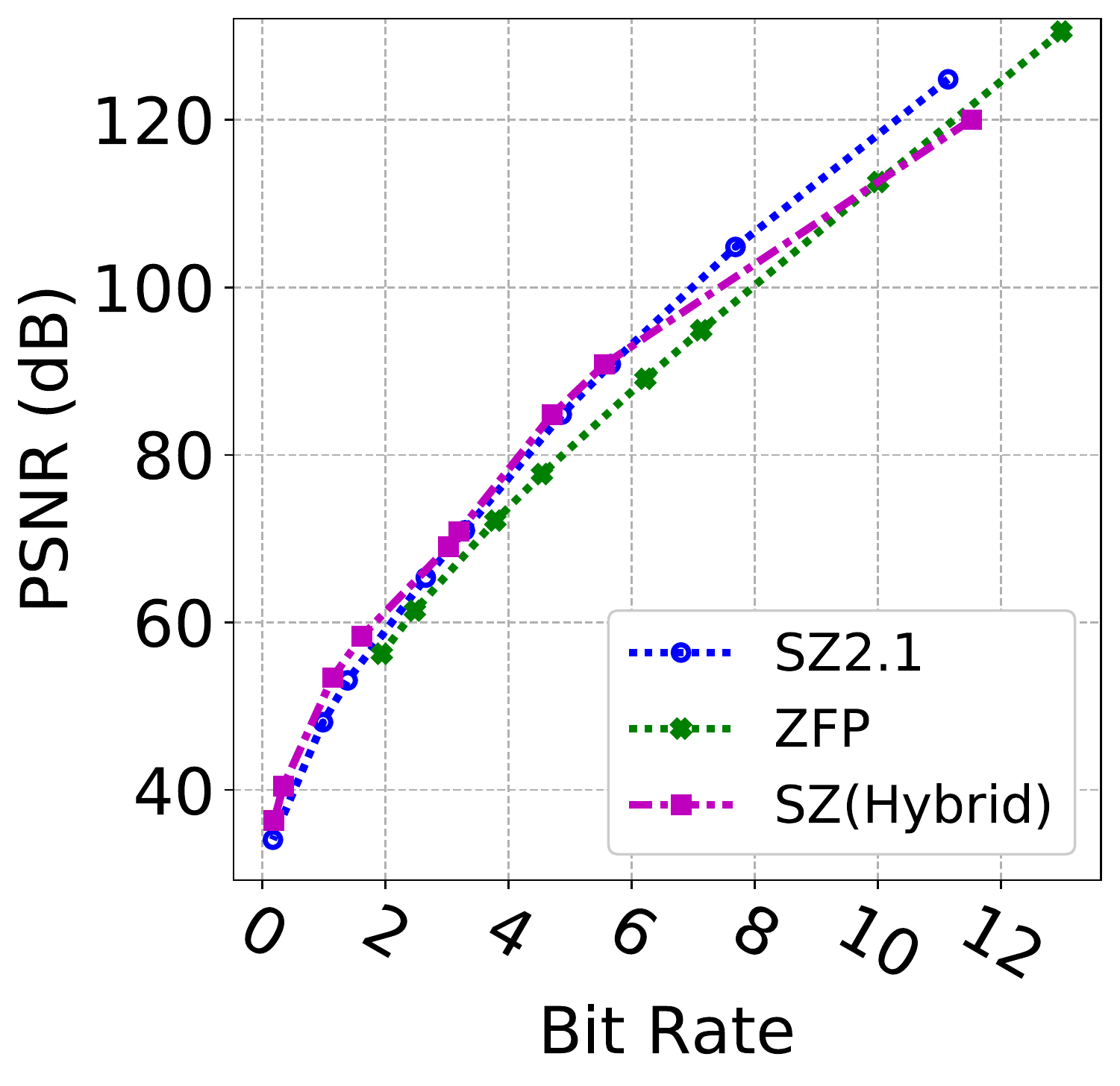}
}}
\hspace{-4mm}
\subfigure[{V}]{
\raisebox{-2mm}{\includegraphics[width=4.45cm]{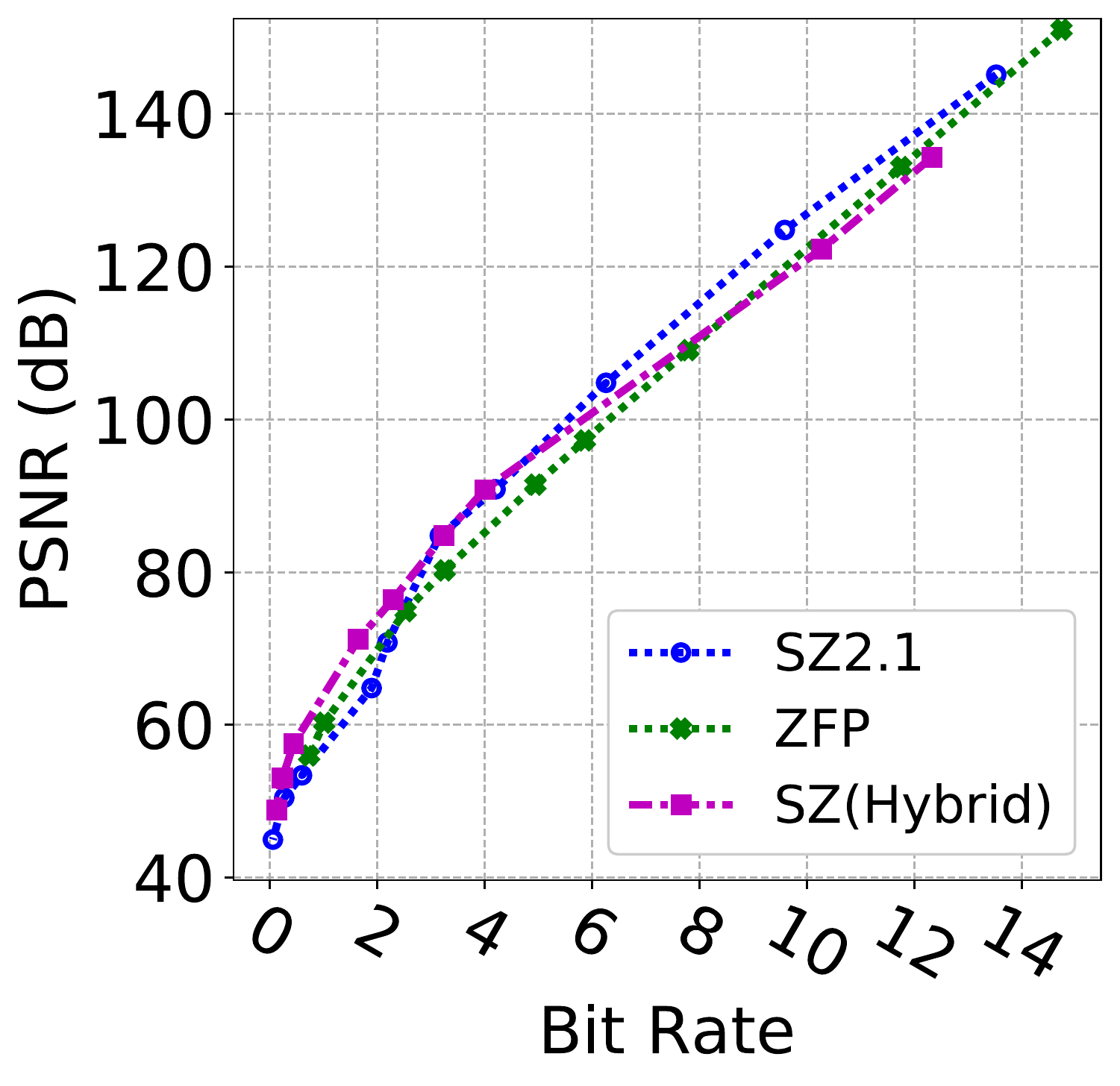}
}}
\hspace{-5mm}
\subfigure[{W}]{
\raisebox{-2mm}{\includegraphics[width=4.45cm]{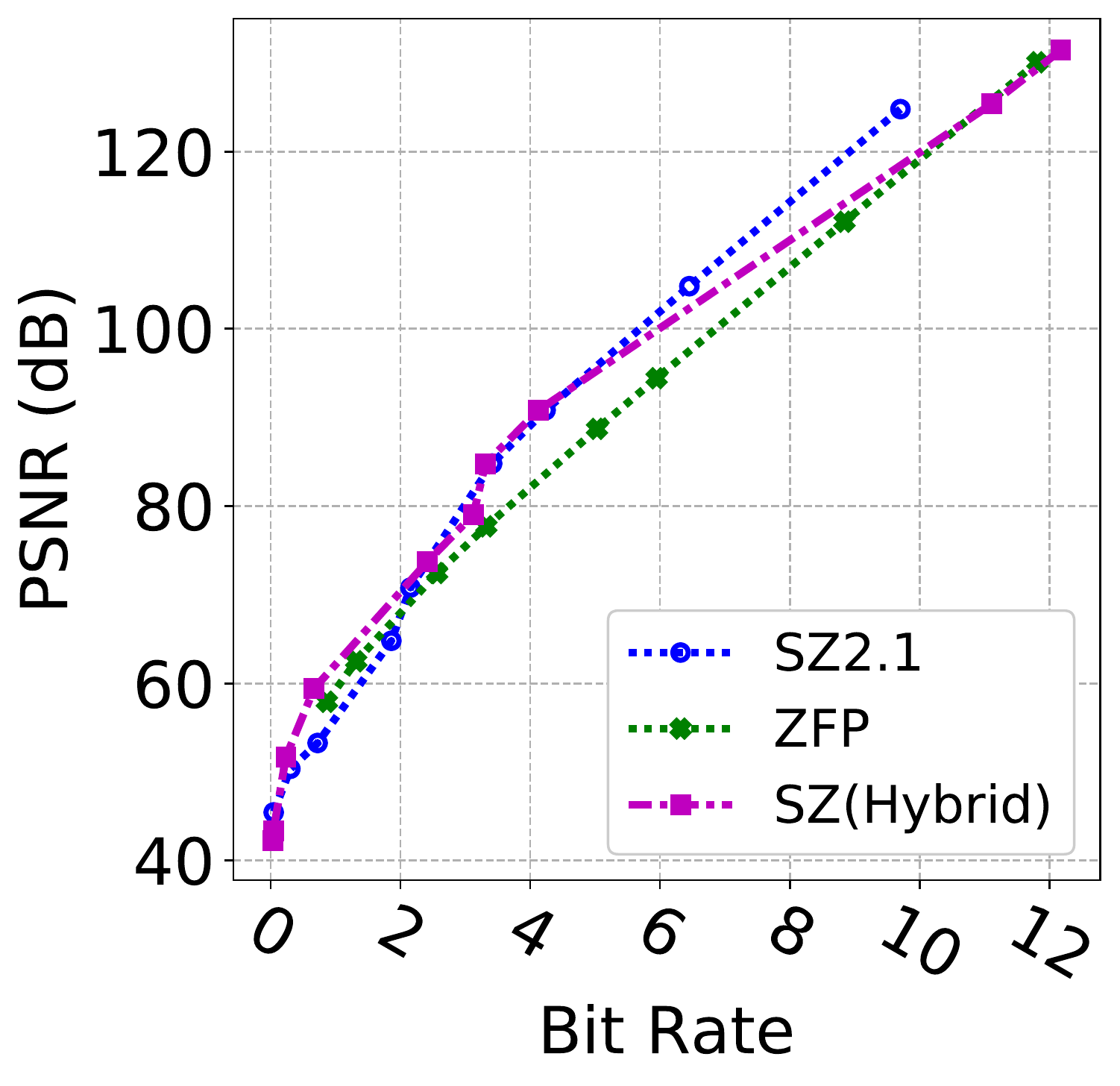}
}}
\hspace{-6mm}

%\vspace{-2mm}
% }
\caption{Rate Distortion (Hurricane-ISABEL)}
\label{fig:ratedistortion-hurricane}
\end{figure*}

\begin{figure*}[ht] \centering
% \hspace{-6mm}
% \subfigure[{CESM-ATM}]{
% \includegraphics[width=4.4cm]{figures/rate-distortion/cesm-eps-converted-to.pdf}
% }
% \hspace{-5mm}
% \subfigure[{Hurricane}]{
% \includegraphics[width=4.4cm]{figures/rate-distortion/loghurricane-eps-converted-to.pdf}
% }
% \hspace{-8mm}

%\hspace{-6mm}
\subfigure[{NYX}]{
\raisebox{-2mm}{\includegraphics[width=4.5cm]{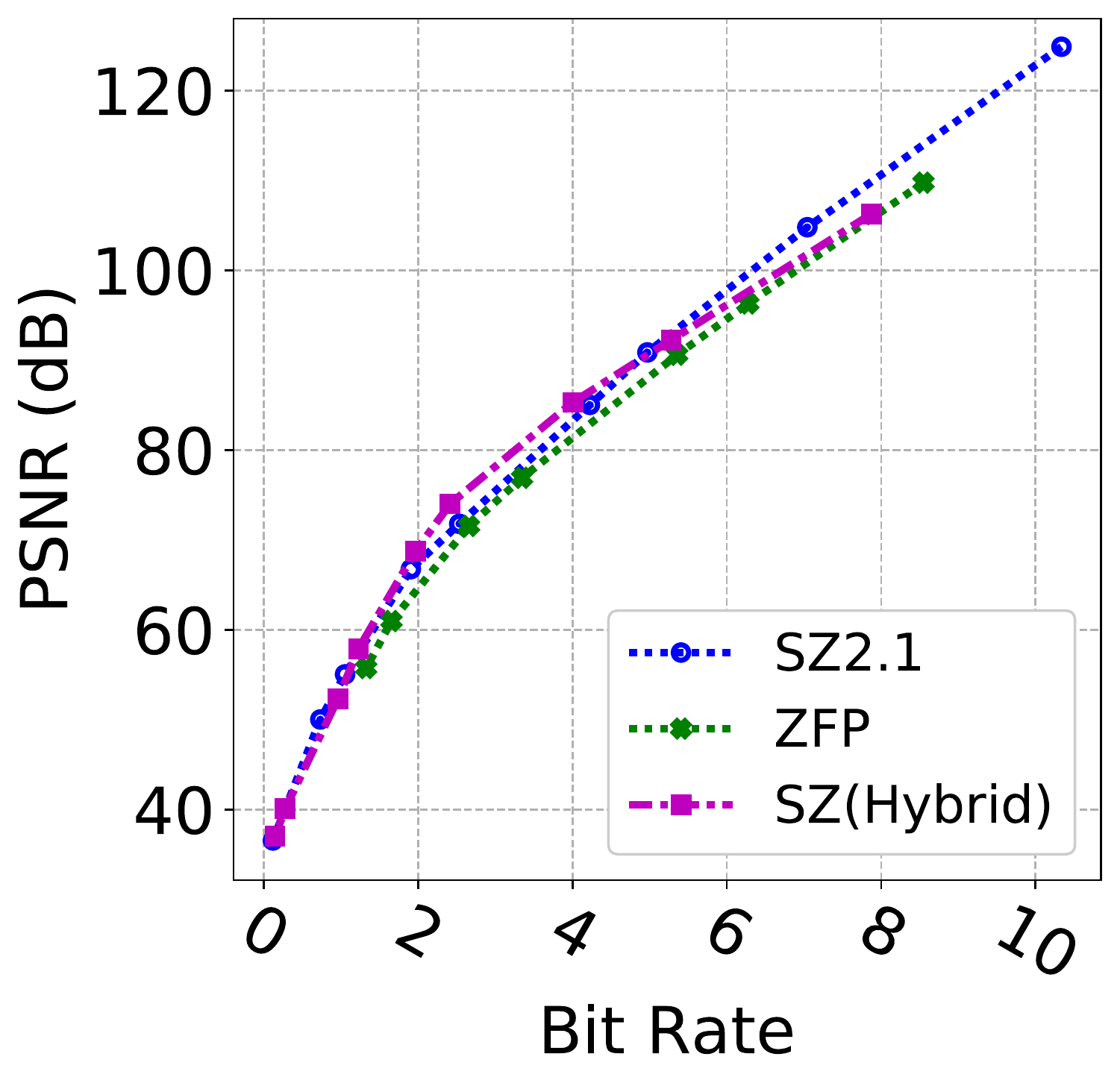}}
}
\hspace{-5mm}
\subfigure[{SCALE-LETKF}]{
\raisebox{-2mm}{\includegraphics[width=4.6cm]{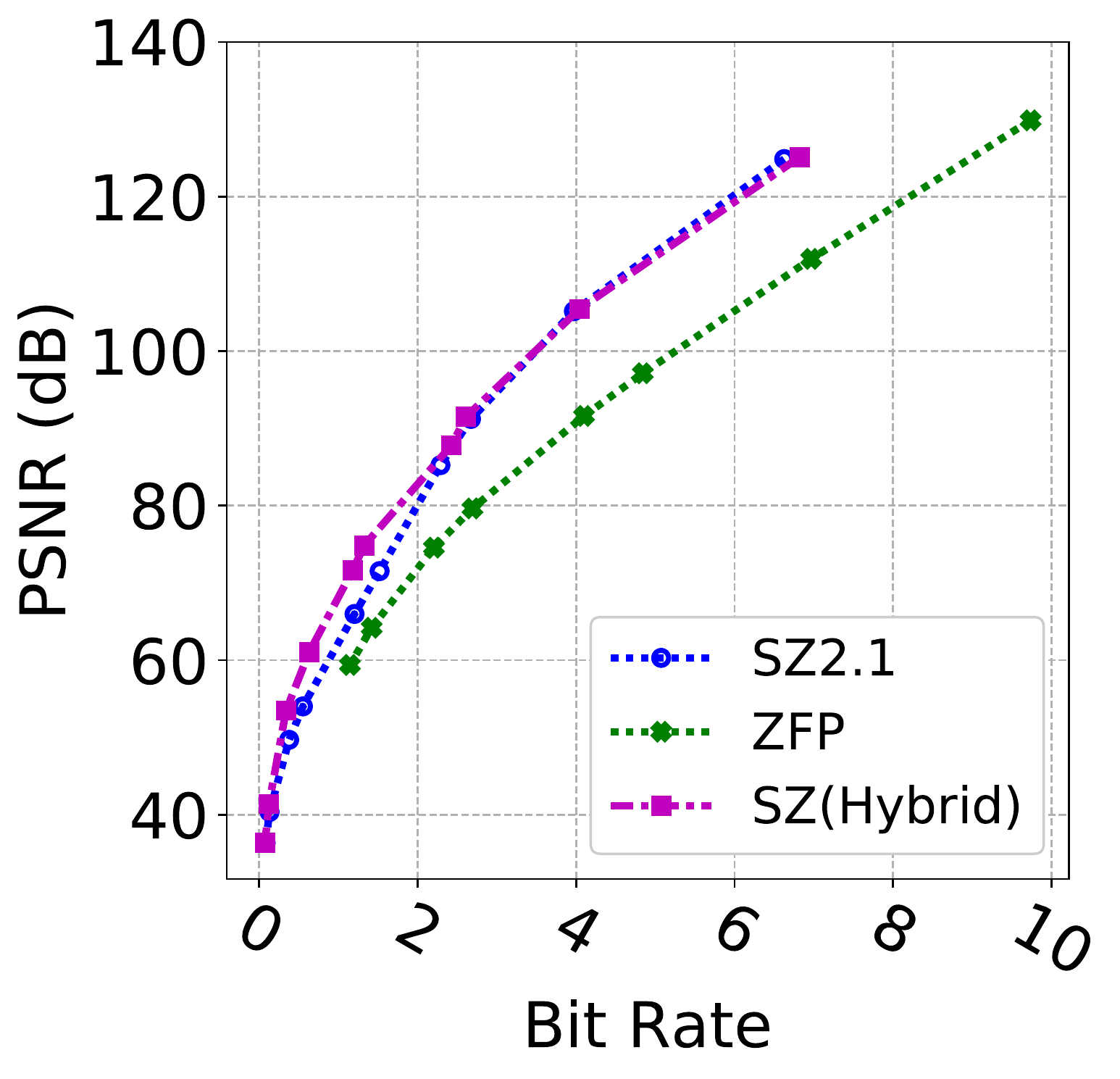}}
}
\hspace{-2mm}
% \vspace{-3mm}
\hspace{-6mm}
\subfigure[{QMCPack}]{
\raisebox{-2mm}{\includegraphics[width=4.5cm]{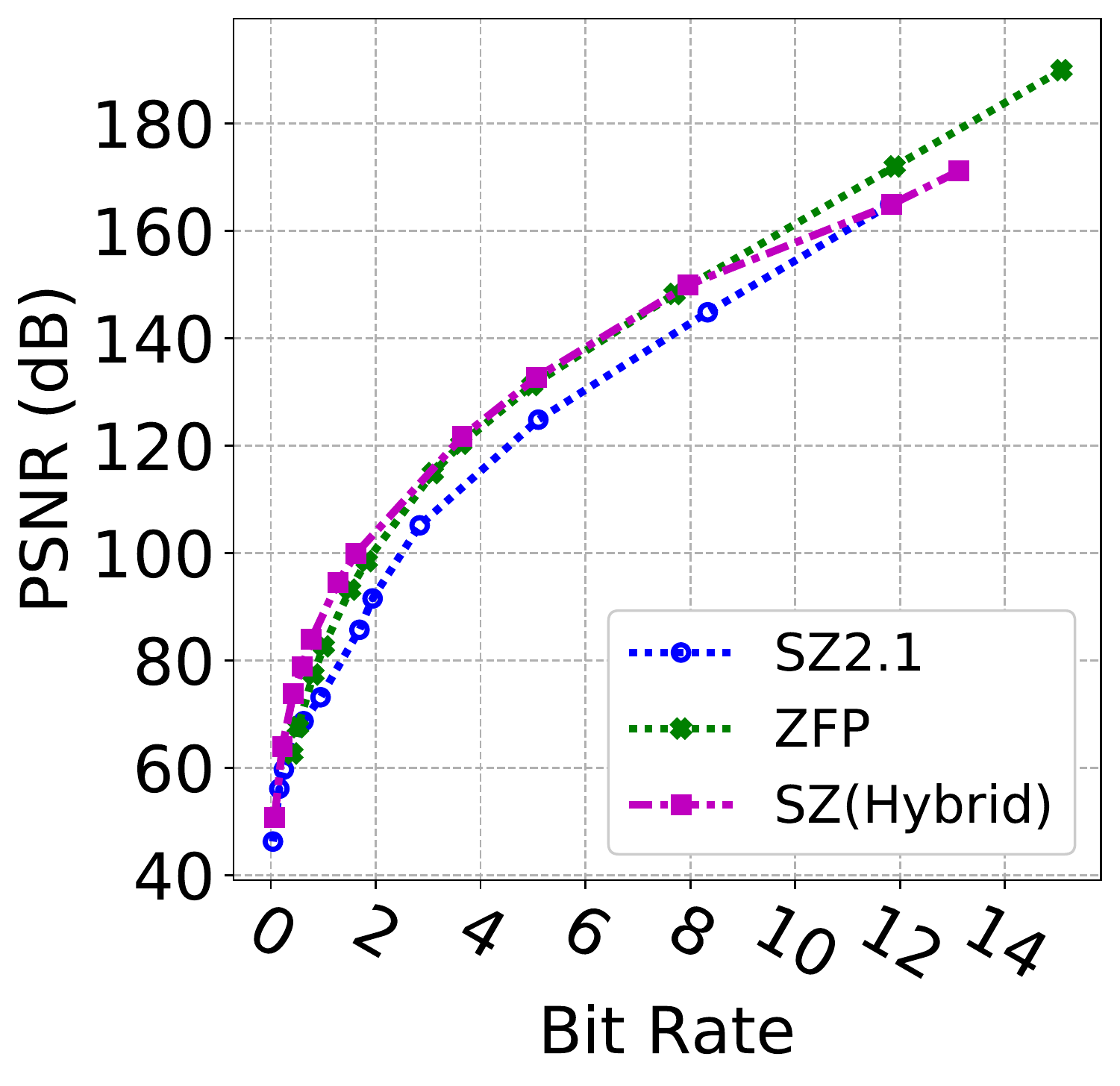}}
}
\hspace{-5mm}
\subfigure[{Hurricane-ISABEL}]{
\raisebox{-2mm}{\includegraphics[width=4.5cm]{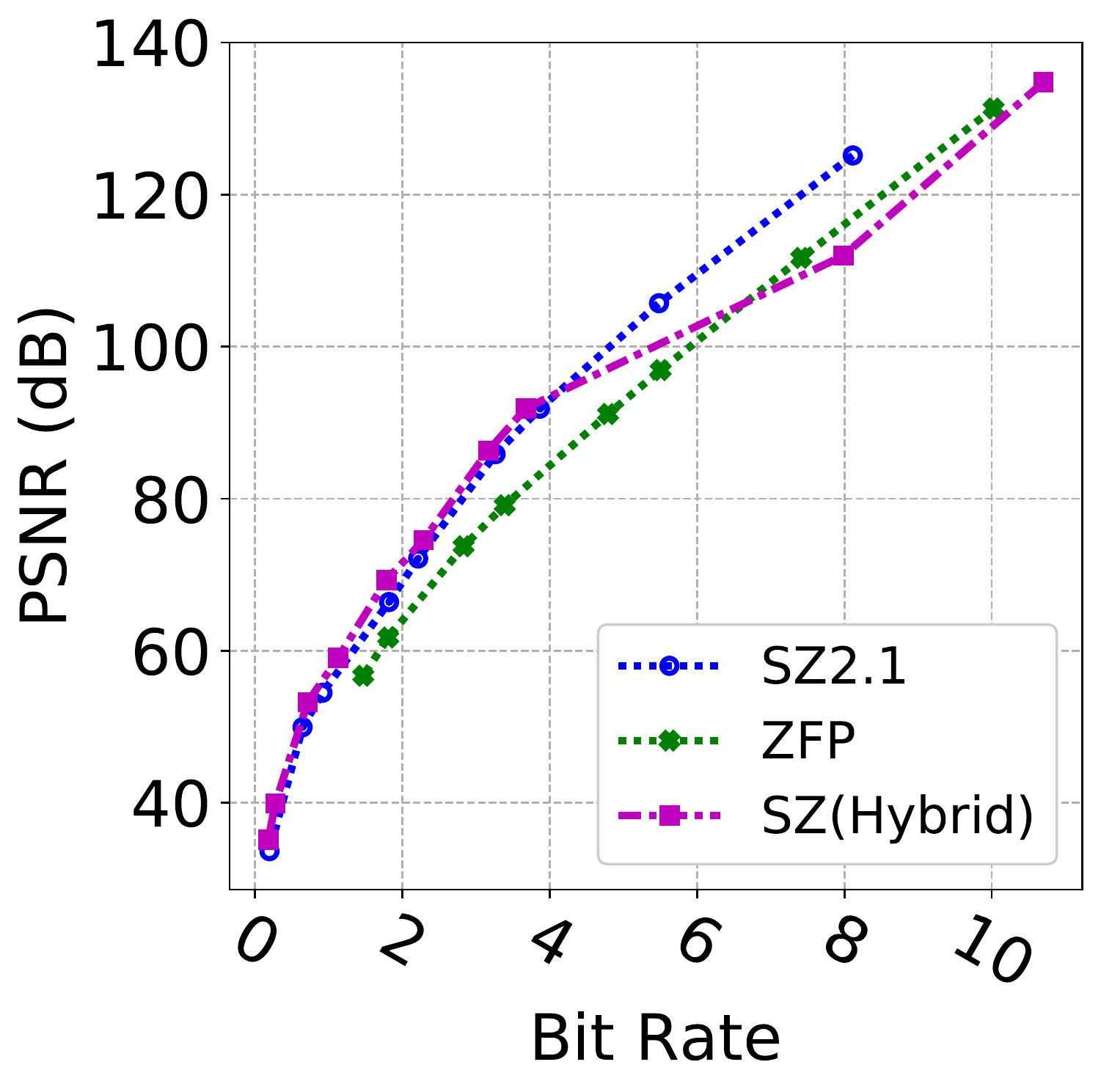}
}}
\hspace{-2mm}

% \vspace{-2mm}
\caption{Rate Distortion (Four Applications)}
\label{fig:ratedistortion-4data}
\end{figure*}

Figure \ref{fig:ratedistortion-hurricane} presents the rate-distortion (i.e., bit-rate versus data distortion) of four typical fields in Hurricane-ISABEL simulation, and Fig. \ref{fig:ratedistortion-4data} presents the rate-distortion for four applications. Bit-rate represents the average number of bits used per data value  after the compression, and data distortion is evaluated using PSNR (the higher the better). 
\textbf{Takeaway 3: Rate-distortion.} When the bit rates are relatively small, SZ(Hybrid) exhibits the best rate-distortion on all four fields of Hurricane-ISABEL and has the best overall rate-distortion on all four applications. On the other hand, SZ is better than SZ(Hybrid) and ZFP when the bit rates are relatively large. For example, on CLOUD field, SZ(Hybrid) has the highest PSNR in the range of [0,2.6], and SZ has the highest PSNR when the bit rate is larger than 2.6.

\begin{figure}[ht] \centering
\hspace{-6mm}
\subfigure[{CESM-ATM(CLOUD)}]{
\includegraphics[width=4.37cm]{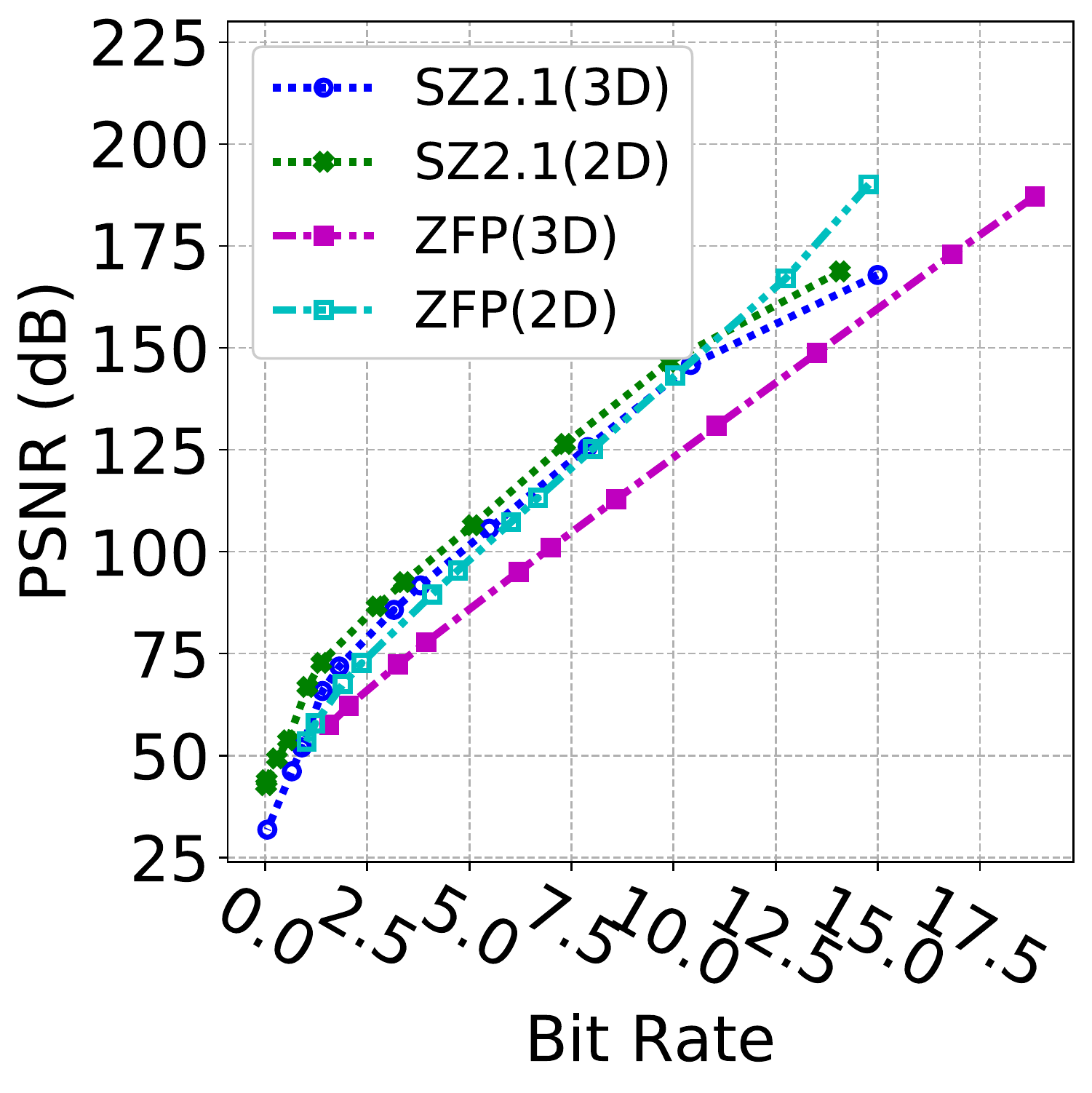}
}
\hspace{-5mm}
\subfigure[{EXAFEL(calibrated)}]{
\includegraphics[width=4.37cm]{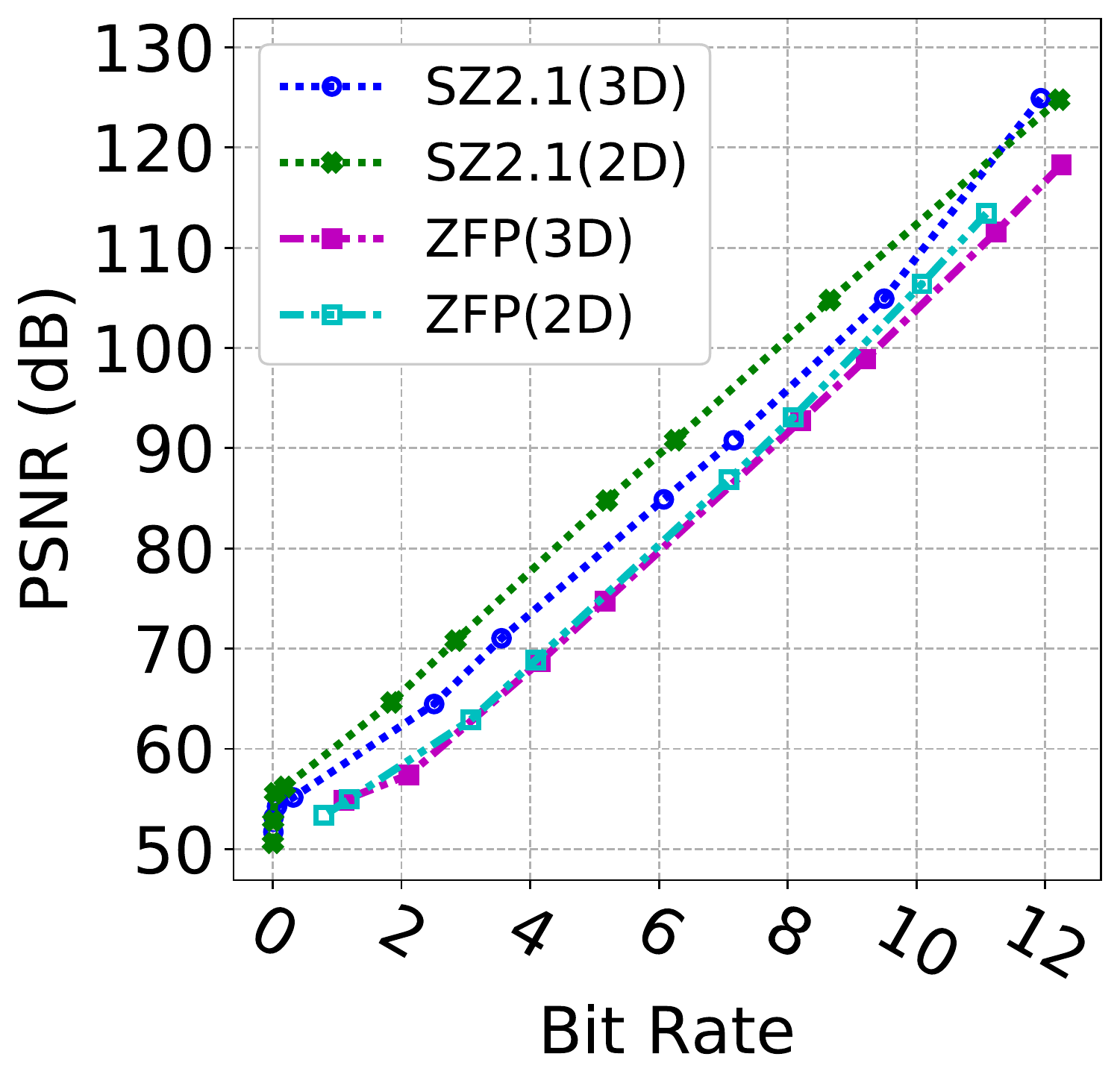}
}
\hspace{-2mm}

% \vspace{-3mm}
\caption{Rate Distortion of CESM and EXAFEL}
\label{fig:ratedistortion-2d3d}
\end{figure}

\textbf{Takeaway 4: Data Dimension.} We observe that treating some 3D datasets including CESM-ATM(CLOUD) and EXAFEL(calibrated) as 2D may improve the compression quality, as demonstrated by rate-distortion in Fig. \ref{fig:ratedistortion-2d3d}. The reason is that CESM-ATM(CLOUD) is  composed of 26 2D arrays ($1800\times3600$ each), and the data is not smooth/consecutive in the third dimension. EXAFEL(calibrated) is composed of 10 3D images ($32\times185\times388$ per image), and each 3D image is combined with 32 2D images captured from separate panels. The non smoothness of data in separate panels makes it better to compress the data as 2D.
SZ(Hybrid) is not included in Fig. \ref{fig:ratedistortion-2d3d} because it does not support compression in 2D mode.

\begin{table*}[ht]
\caption{Compression Quality (Compression Ratio, PSNR and SSIM)} \centering
% \vspace{-2mm}
\scriptsize
\resizebox{2\columnwidth}{!}{        
\begin{tabular}{|c||c|c|c|c|c|c|c|c|c|c|c|c|c|}
\hline \multirow{2}{*}{\textbf{Dataset}} & \multicolumn{4}{|c|}{\textbf{SZ}} & \multicolumn{4}{|c|}{\textbf{ZFP}} &
\multicolumn{5}{|c|}{\textbf{Compression Ratio of Lossless Compressors}} \\
\cline{2-14} & \textbf{CR} & \textbf{PSNR} & \textbf{SSIM(1D)} & \textbf{SSIM(2D)} & \textbf{CR} & \textbf{PSNR} & \textbf{SSIM(1D)} &\textbf{SSIM(2D)} & \textbf{ZSTD} & \textbf{C-Blosc2} & \textbf{FPZIP} & \textbf{FPC} & \textbf{ZFP} \\
\hline QMCPack(einspline) & 14.35 & 96 & $>$0.9999995 & 0.9837 & 14.6 & 104.2 & $>$0.9999995 & 0.985 & 1.20 & 1.01 & 1.75 & 1.09 & 2.21 \\
\hline Hurricane-ISABLE(P) & 58.3 & 68.838 & 0.999989 & 0.9963 & 44.7 & 64.841 & 0.997443 & 0.9877 & 1.15 & 1.00 & 2.11 & 1.10 & 1.64 \\
\hline CESM-ATM(CLOUD) & 30.75 & 66.8 & 0.999988 & 0.999999 & 31.5 & 53.4 & 0.999734 & 0.9815 & 1.66 & 1.40 & 2.32 & 1.54 & 1.58\\
\hline EXAFEL (calibrated) & 23.4 & 67.5 & 0.999999 & 0.999999 & 4.73 & 64.77 & 0.999999 & 0.99998 & 1.96 & 1.11 & 1.11 & 1.00 & 3.29 \\
\hline
\end{tabular}
}
\label{tab:psnrssim}
% \vspace{-5mm}
\end{table*}

Table \ref{tab:psnrssim} presents the PSNR and SSIM (in terms of both 1D and 2D) for the four datasets with different compressors, by tuning the compression ratios to the similar level (except for EXAFEL because the compression ratios of ZFP are always lower than 8:1). \textbf{Takeaway 5: SSIM and PSNR.} It is observed that 1D SSIM is always very close to 1 for any compressor, while 2D SSIM and PSNR can show the discrepancies of compression results more clearly on different compressors. The reason 1D SSIM always approaches to 1 is that the mean, standard deviation and covariance are always very similar between the original data and decompressed data. Some recent studies \cite{vis-ssim, ssim2} on visualization shows that 2D SSIM could be more accurate than PSNR in some cases, so we suggest to use both PSNR and 2D SSIM in the evaluation of lossy compression quality. 

Table \ref{tab:psnrssim} also contains compression ratios of five state-of-the-art lossless compressors. We include ZFP since it has a lossless mode besides the lossy mode. 
\textbf{Takeaway 6: Lossless versus Lossy.} Lossless compressors generally have very low compression ratios on scientific datasets. Their compression ratios are in the range of 1$\sim$3 which is far from desired levels to solve the storage and I/O bottleneck problem. Lossy compressors, on the other hand, can reach 20$\times$ higher compression ratio than lossless compressors do, with acceptable data fidelity for post-analysis based on user-specified error bounds. As a result, lossy compressors are more suitable for scientific data compression scenarios. 
\section{Conclusion and Future Work}
\label{sec:con}

In this paper, we release SDRBench, a scientific data reduction benchmark to help compression users and developers assess lossy compressors fairly and conveniently. SDRBench contains scientific datasets, lossy compressor assessment metrics, and state-of-the-art lossy compressors. The 10+ scientific datasets in SDRBench are from different domains including climate simulation, cosmological n-body simulation, turbulence simulation, and others.
We also present the evaluation results using SDRBench and summarize six valuable takeaways to help developers and users to have a better understanding of lossy compressors.
\begin{itemize}
    \item \textbf{Takeaway 1: Compression Error.} Lossy compressors have different compression error distributions. In addition, ZFP tends to over preserve the compression precision.
    \item \textbf{Takeaway 2: Compression Speed.} Lossy compressors have varied compression speed. For example, ZFP is about $10\% \sim 100\%$ faster than SZ.
    \item \textbf{Takeaway 3: Rate-distortion.} Currently, no  lossy compressor can always outperform the others in terms of rate-distortion. 
    \item \textbf{Takeaway 4: Data Dimension.} For some 3D data such as the CLOUD field of CESM-ATM, treating them as 2D data may improve the compression quality.
    \item \textbf{Takeaway 5: SSIM and PSNR.} 2D SSIM is better than 1D to show the discrepancies of compression results. We suggest use 2D SSIM and PSNR in the evaluation of lossy compression quality.  
    \item \textbf{Takeaway 6: Lossless versus Lossy.} Lossless compressors have very low compression ratio (usually 1$\sim$3) on scientific datasets, while lossy compressors can achieve $20\times$ higher compression ratio on the same dataset than lossless compressors.
\end{itemize}

In the future, we will include more datasets in different scientific domains and more lossy compressors in the benchmark.

% we will improve our developed compression assessment tool - Z-checker by enabling it to support more evaluation metrics based on users' demand.

\section{Acknowledgments}
%\footnotesize
This research was supported by the Exascale Computing Project (ECP), Project Number: 17-SC-20-SC, a collaborative effort of two DOE organizations – the Office of Science and the National Nuclear Security Administration, responsible for the planning and preparation of a capable exascale ecosystem, including software, applications, hardware, advanced system engineering and early testbed platforms, to support the nation’s exascale computing imperative. The material was supported by the U.S. Department of Energy, Office of Science, under contract DE-AC02-06CH11357, and supported by the National Science Foundation under Grant No. 1619253. We acknowledge the computing resources provided on Bebop, which is operated by the Laboratory Computing Resource Center at Argonne National Laboratory.

\bibliographystyle{IEEEtran}
\bibliography{bib/refs}

% The submitted manuscript has been created by UChicago Argonne, LLC, Operator of Argonne National Laboratory (``Argonne'').  Argonne, a U.S. Department of Energy Office of Science laboratory, is operated under Contract No. DE-AC02-06CH11357.  The U.S. Government retains for itself, and others acting on its behalf, a paid-up nonexclusive, irrevocable worldwide license in said article to reproduce, prepare derivative works, distribute copies to the public, and perform publicly and display publicly, by or on behalf of the Government.

\end{document}